\documentclass[10pt,conference]{IEEEtran}
\IEEEoverridecommandlockouts
\usepackage{cite}
\usepackage{amsmath,amssymb,amsfonts}
\usepackage{graphicx}
\usepackage[ruled,linesnumbered]{algorithm2e}
\usepackage{textcomp}
\usepackage{subcaption}
\usepackage{xcolor}
\usepackage{xspace}
\usepackage{url}
\usepackage{booktabs}
\usepackage{makecell}
\usepackage{multirow}
\usepackage{threeparttable}

\def\BibTeX{{\rm B\kern-.05em{\sc i\kern-.025em b}\kern-.08em
    T\kern-.1667em\lower.7ex\hbox{E}\kern-.125emX}}
\begin{document}

\newcommand{\argmin}{\mathop{\mathrm{argmin}}}   
\newcommand{\fixme}[1]{\textcolor{red}{\textbf{fixme:}#1}}
\newcommand{\sy}[1]{\textcolor{red}{\textbf{SY:}#1}}
\newcommand{\rf}[1]{\textcolor{red}{\textbf{RF:}#1}}
\newcommand{\jk}[1]{\textcolor{red}{\textbf{JK:}#1}}
\newcommand{\name}{SADL\xspace}

\title{
Guiding Deep Learning System Testing using Surprise Adequacy
}

\author{\IEEEauthorblockN{Jinhan Kim}
\IEEEauthorblockA{\textit{School of Computing}\\
\textit{KAIST}\\
Daejeon, Republic of Korea\\
jinhankim@kaist.ac.kr}
\and
\IEEEauthorblockN{Robert Feldt}
\IEEEauthorblockA{\textit{Dept. of Computer Science and Engineering} \\
\textit{Chalmers University}\\
Gothenburg, Sweden \\
robert.feldt@chalmers.se}
\and
\IEEEauthorblockN{Shin Yoo}
\IEEEauthorblockA{\textit{School of Computing} \\
\textit{KAIST}\\
Daejeon, Republic of Korea \\
shin.yoo@kaist.ac.kr}
}

\maketitle

\begin{abstract}
Deep Learning (DL) systems are rapidly being adopted in safety and security 
critical domains, urgently calling for ways to test their correctness and 
robustness. Testing of DL systems has traditionally relied on manual 
collection and labelling of data. Recently, a number of coverage criteria 
based on neuron activation values have been proposed. These criteria 
essentially count the number of neurons whose activation during the execution 
of a DL system satisfied certain properties, such as being above predefined 
thresholds. However, existing coverage criteria are not sufficiently fine 
grained to capture subtle behaviours exhibited by DL systems. Moreover, 
evaluations have focused on showing correlation between adversarial examples 
and proposed criteria rather than evaluating and guiding their use for actual 
testing of DL systems. We propose a novel test adequacy criterion for testing 
of DL systems, called Surprise Adequacy for Deep Learning Systems (SADL), 
which is based on the behaviour of DL systems with respect to their training 
data. We measure the surprise of an input as the difference in DL system's 
behaviour between the input and the training data (i.e., what was learnt 
during training), and subsequently develop this as an adequacy criterion: a 
good test input should be sufficiently but not overtly surprising compared to 
training data. Empirical evaluation using a range of DL systems from simple 
image classifiers to autonomous driving car platforms shows that systematic 
sampling of inputs based on their surprise can improve classification accuracy 
of DL systems against adversarial examples by up to 77.5\% via retraining.

\end{abstract}


\section{Introduction}
\label{sec:introduction}

Deep Learning (DL)~\cite{LeCun2015ef} systems have achieved significant 
progress in many domains including image recognition~\cite{Krizhevsky2012hh,
Farabet2013bc,Szegedy2015pt}, speech recognition~\cite{Hinton2012lg}, and 
machine translation~\cite{Jean2015mw,Sutskever2014tz}. Based on their 
capability to match or even surpass human performance, DL systems are 
increasingly being adopted as part of larger systems in both safety and 
security critical domains such as autonomous driving~\cite{Bojarski2016ak,
chen2015deepdriving}, and malware detection~\cite{cui2018detection}. 

Such adoption of DL systems calls for new challenges, as it is critically 
important that these larger systems are both correct and predictable. Despite 
their impressive experimental performances, DL systems are known to exhibit 
unexpected behaviours under certain circumstances. For example, in a reported 
incident, an autonomous driving vehicle expected another vehicle to yield in 
one of the rarer circumstances, and crashed into the other vehicle when the 
expectation proved incorrect~\cite{google2016uv}. There is an urgent need to 
verify and validate behaviours of DL systems. However, a significant part of 
existing software testing technique is not directly applicable to DL systems. 
Most notably, traditional white-box testing techniques that aim to increase 
structural coverage~\cite{Ammann2016uu} is not very useful for DL systems, as 
their behaviour is not explicitly encoded in their control flow structures.

A number of novel approaches towards testing and verification of DL systems 
have been recently proposed to fill in the gap~\cite{Huang2017kx,Pei2017qy,
Tian2018zn,Ma2018ny}. Most of these techniques share two assumptions. The 
first assumption is essentially a generalisation of the essence of metamorphic 
testing~\cite{Chen:2004th}: if two inputs to a DL system are \emph{similar} 
with respect to some human sense, the outputs should also be similar. For 
example, DeepTest~\cite{Tian2018zn} checks whether an autonomous driving 
system behaves in the same way when the input image is transformed as if the 
same scene is under a different weather condition. The second assumption, also 
based in more traditional software testing results~\cite{Feldt:2016if}, is 
that the more diverse a set of input is, the more effective testing of a DL 
system one can perform. For example, DeepXplore~\cite{Pei2017qy} presented the 
Neuron Coverage (the ratio of neurons whose activation values were above a 
predefined threshold) as the measure of diversity of neuron behaviour, and 
subsequently showed that inputs violating the first assumption will also 
increase the neuron coverage.

While the recently introduced techniques have made significant advances over 
manual \emph{ad hoc} testing of DL systems, there is a major limitation.
The coverage criteria proposed so far are not sufficiently fine grained,
in a sense that all of them simply count neurons whose activation values 
satisfy certain conditions. While this aggregation by counting does allow the
tester to quantify the test effectiveness of a given input \texttt{set}, it
conveys little information about individual inputs. For example, it is not
immediately clear when an input with higher NC should be considered 
\emph{better} than another with lower NC, and why: certain inputs may
naturally activate more neurons above the threshold than others, and vice 
versa. Another example is the $k$-Multisection Neuron 
Coverage~\cite{Ma2018ny}, which partitions the ranges of activation values of 
neurons, observed during training, into $k$ buckets, and count the number of 
total buckets covered by a set of inputs. When measured for a single input, 
the coverage will be either $\frac{1}{k}$ if the input activates each neuron 
with a value from one of the $k$ buckets, or smaller than that if some neurons 
activate outside the range observed during training. Again, the information 
about how far such activations go beyond observed range is lost during 
aggregation, making it hard to evaluate the relative value of each input. 
For a test adequacy criterion to be practically useful, it should be able
to guide the selection of individual inputs, eventually resulting in 
improvements of the accuracy of the DL system under investigation.

To overcome these limitations, we propose a new test adequacy for DL systems, 
called Surprise Adequacy for DL systems (\name). %
Intuitively, a good test 
input set for a DL system should be systematically diversified to include 
inputs ranging from those similar to training data to those significantly 
different and adversarial.\footnote{Experiments show benefits of diversity 
for general testing~\cite{Feldt:2016if} and benefits of a `scale of distances' 
of test inputs for robustness testing introduced in~\cite{Poulding2017aa}.}
At individual input granularity, \name measures how 
\emph{surprising} the input is to a DL system with respect to the data the 
system was trained with: the actual measure of surprise can be either based on the 
likelihood of the system having seen a similar input during training (here with 
respect to probability density distributions extrapolated from the training process 
using kernel density estimation~\cite{wand1994kernel}), or the distance between vectors
representing the neuron activation traces of the given input and the training data (here simply using Euclidean distance).
Subsequently, the Surprise Adequacy (SA) of a set of test inputs is measured by 
the range of individual surprise values the set covers. We show that \name is 
sufficiently fine rained by training adversarial example classifiers based on 
\name values that can produce higher accuracy compared to the state of the 
art. We also show that sampling inputs according to \name for retraining DL 
systems can result in higher accuracy, thus showing that \name is an 
independent variable that can positively affect the effectiveness of DL system 
testing.

The technical contributions of this paper are as follows:

\begin{itemize}
\item We propose \name, a fine grained test adequacy metric that measures the 
surprise of an input, i.e., the difference in the behaviour of a DL system 
between a given input and the training data. Two concrete instances of \name 
are proposed based on different ways to quantify surprise. Both are shown
to be correlated with existing coverage criteria for DL systems.

\item We show that \name is 
sufficiently fine grained in capturing the behaviour of DL systems by training 
a highly accurate adversarial example classifier.
Our adversarial example classifier shows as much as 100\% and 94.53\% ROC-AUC score
when applied to MNIST~\cite{LeCun2010vg} and CIFAR-10~\cite{Krizhevsky2014nu} 
dataset, respectively.

\item We show that \name metrics can be used to sample effective test input 
sets. When retraining DL systems using additional adversarial examples, 
sampling additional inputs with broader SA values can improve the accuracy after
retraining by up to 77.5\%.

\item We undertake all our experiments using publicly available DL systems 
ranging from small benchmarks (MNIST and CIFAR-10) to a large system for 
autonomous driving vehicles
(Dave-2~\cite{Bojarski2016ak} and Chauffeur~\cite{Chauffeur}). 
\end{itemize}

The remaining of this paper is organised as follows. Section~\ref{sec:sadl} 
introduces Surprise Adequacy for DL systems, \name: two variants of \name are 
presented along with algorithms that measure them. Section~\ref{sec:rqs} 
sets out the research questions and Section~\ref{sec:setup} describes the experimental set-up of the empirical evaluations.
Section~\ref{sec:result} presents the results from empirical evaluations. 
Section~\ref{sec:threats} addresses threats to validity. 
Section~\ref{sec:relatedwork} presents related work, and Section~\ref{sec:conclusion} concludes.

\section{Surprise Adequacy for Deep Learning Systems}
\label{sec:sadl}

All existing test adequacy criteria for DL systems aim to measure the 
diversity of an input set. Neuron Coverage (NC)~\cite{Pei2017qy} posits that 
the higher the number of neurons that are activated above a predefined 
threshold, the more diverse input the DL system has been executed with. 
DeepGauge~\cite{Ma2018ny} proposed a range of finer grained adequacy criteria 
including $k$-Multisection Neuron Coverage, which measures the ratio of 
activation value \emph{buckets} that have been covered across all neurons, and 
Neuron Boundary Coverage, which measures the ratio of neurons that are 
activated beyond the ranges observed during training. 

We argue that diversity in testing of DL systems is more %
meaningful when it is 
measured with respect to the training data, as DL systems are likely to be 
more error prone for inputs that are unfamiliar, i.e., diverse. 
Furthermore, 
while neuron activation above thresholds, or beyond observed ranges, may be 
closely related to diversity of the given input, they do not measure to what 
degree the activations of the network for one input differs from the 
activations for another input.
They are fundamentally discretisations and do not utilize the fact that neuron 
activations are continuous quantities.
In contrast, our aim is to define an adequacy criterion that quantitatively 
measures behavioural differences observed in a given set of inputs, relative 
to the training data.

\subsection{Activation Trace and Surprise Adequacy}
\label{sec:activation_traces}

Let $\mathbf{N} = \{n_1, n_2, \ldots\}$ be a set of neurons that constitutes a 
DL system $\mathbf{D}$, and let $X = \{x_1, x_2, \ldots\}$ be a set of inputs. 
We 
denote the activation value of a single neuron $n$ with respect to an input $x$
as $\alpha_n(x)$. For an ordered (sub)set of neurons, 
let $N \subseteq \mathbf{N}$, $\alpha_N(x)$ denote a vector of activation 
values, each element corresponding to an individual neuron in $N$: the 
cardinality of $\alpha_N(x)$ is equal to $|N|$. We call $\alpha_N(x)$ the 
Activation Trace (AT) of $x$ over neurons in $N$. 
Similarly, let $A_N(X)$ be 
a set of activation traces, observed over neurons in $N$, for a set of inputs $
X$: $A_N(X) = \{\alpha_N(x) \mid x \in X\}$. We note that the activation trace is trivially
available after each execution of the network for a given input. 

Since behaviours of DL systems are driven along the data-flow and not 
control-flow, we assume that activation traces observed over all $\mathbf{N}$ 
with respect to $X$, $A_\mathbf{N}(X)$, fully captures 
the behaviours of the DL system under investigation when executed using 
$X$.\footnote{For the sake of simplicity, we assume that it is 
possible to get the complete activation traces from all the neurons in a DL 
system. For network architectures with loops, such as Recurrent Neural Nets 
(RNNs)~\cite{Hochreiter1997aa}, it is possible to \emph{unroll}
the loops up to a predefined bound~\cite{Tian2018zn}.} 

Surprise Adequacy (SA) aims to measure the relative novelty (i.e., surprise) of
a given new input with respect to the inputs used for training. 
Given a 
training set $\mathbf{T}$, we first compute $A_\mathbf{N}(\mathbf{T})$
by recording activation values of all neurons using every input in the training
data set. Subsequently, given a new input $x$, we measure how surprising $x$ 
is when compared to $\mathbf{T}$ by comparing the activation trace of $x$ to 
$A_\mathbf{N}(\mathbf{T})$. This quantitative similarity measure is called
Surprise Adequacy (SA). We introduce two variants of SA, each with different
way of measuring the similarity between $x$ and 
$A_\mathbf{N}(\mathbf{T})$.\footnote{However, the main idea is general and 
other, specific variants would result if using other similarity functions.}

Note that certain types of DL tasks allow us to focus on parts of
the training set $\mathbf{T}$ to get more precise and meaningful measurement of SA. 
For example, suppose we are testing a classifier with a new input $x$, which is 
classified by the DL system under investigation as the class $c$. In this 
case, the surprise of $x$ is more meaningfully measured against $A_\mathbf{N}(
T_c)$, in which $T_c$ is the subset of $\mathbf{T}$ where members are 
classified as $c$. Basically, the input might be surprising as an example of 
class $c$ even if not surprising in relation to the full set of training examples.

\subsection{Likelihood-based Surprise Adequacy}
\label{sec:likelihood_sa}

Kernel Density Estimation (KDE)~\cite{wand1994kernel} is a way of estimating 
the probability density function of a given random variable. The resulting
density function allows the estimation of relative likelihood of a specific
value of the random variable. Likelihood-based 
SA (LSA) uses KDE to estimate the probability density of each activation value 
in $A_\mathbf{N}(\mathbf{T})$, and obtains the surprise of a new input with 
respect to the estimated density. This is an extension of existing work that
uses KDE to detect adversarial examples~\cite{Feinman2017aa}.
To reduce dimensionality and computational 
cost, we only consider the neurons in a selected layer $N_{L} \subseteq 
\mathbf{N}$, which yields a set of activation traces, $A_{N_{L}}(\mathbf{X})$. 
To further reduce the computational cost, we filter out neurons whose activation 
values show variance lower than a pre-defined threshold, $t$, as these neurons
will not contribute much information to KDE.
The cardinality of each trace will be $|N_{L}|$. Given a bandwidth matrix $H$ 
and Gaussian kernel function $K$, the activation trace of the new input $x$, 
and $x_i \in \mathbf{T}$, KDE produces density function $\hat{f}$ as follows:

\begin{equation}
    \hat{f}(x) = \frac{1}{|A_{N_{L}}(\mathbf{T})|}\sum_{x_i \in 
    \mathbf{T}}{K_{H}(\alpha_{N_{L}}(x)-\alpha_{N_{L}}(x_i))}
\label{eq:kde}
\end{equation}

Since we want to measure the \emph{surprise} of the input $x$, we need a metric
that increases when probability density decreases (i.e., the input is rarer
compared to the training data), and vice versa (i.e., the input is similar to 
the training data). Adopting common approach of converting probability density 
to a measure of rareness~\cite{tarassenko2005biosign,Luca2014qy},
we define LSA to be the negative of the log of density:

\begin{equation}
    LSA(x) = -log(\hat{f}(x))
\label{eq:ksa}
\end{equation}

Note that extra information about input types can be used to make LSA
more precise. For example, given a DL classifier $\mathbf{D}$, we expect 
inputs that share the same class label will have similar ATs. We can exploit 
this by computing LSA per class, replacing $\mathbf{T}$ with $\{x \in 
\mathbf{T} \mid \mathbf{D}(x) = c\}$ for class $c$. We use per-class 
LSA for DL classifiers in our empirical evaluation.

\begin{figure}[ht]
\centering
\includegraphics[width=65mm]{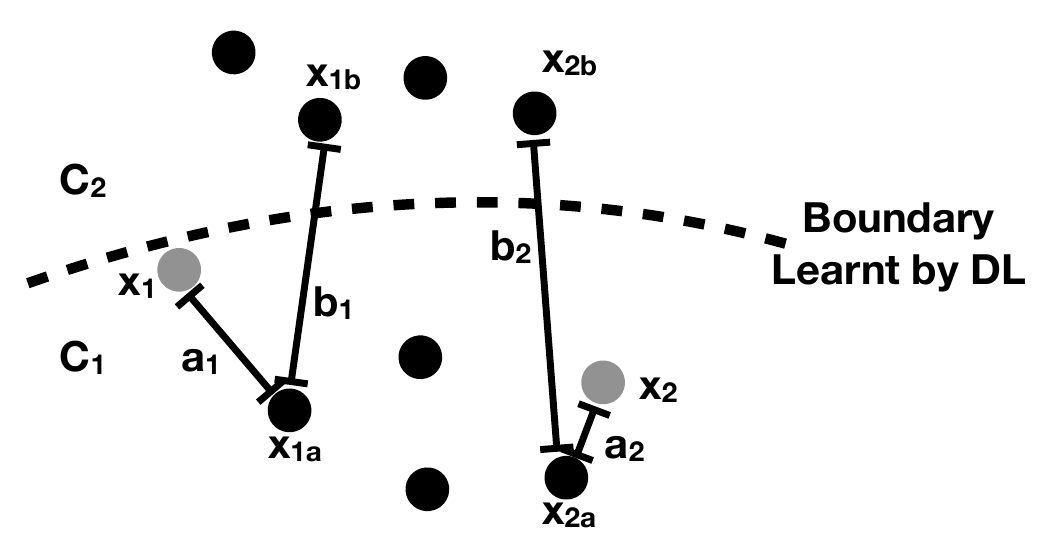}
\caption{An example of Distance-based SA. Black dots represent ATs of training 
data inputs, whereas grey dots represent ATs of new inputs, $x_1$ and 
$x_2$. Compared to distances from $x_{1a}$ and $x_{2a}$ to class 
$c_2$, AT of $x_1$ is farther out from class $c_1$ than that of $x_2$, i.e., 
$\frac{a_1}{b_1} > \frac{a_2}{b_2}$ (see Equations~\ref{eq:dist_a},~\ref{eq:dist_b}, and~\ref{eq:dsa}). Consequently, we decide that $x_1$ is 
more surprising than $x_2$ w.r.t. class $c_1$.\label{fig:dsa_example}}
\end{figure}

\subsection{Distance-based Surprise Adequacy}
\label{sec:distance_sa}

An alternative to LSA is simply to use the distance between ATs
as the measure of surprise. Here, we define Distance-based Surprise Adequacy 
(DSA) using the Euclidean distance between the AT of a new input $x$ and 
ATs observed during training. Being a distance metric,
DSA is ideally suited to exploit the boundaries between inputs, as can be
seen in the classification example in Figure~\ref{fig:dsa_example}. By 
comparing the distances $a_1$ and $a_2$ (i.e., distance between the AT of a 
new input and the reference point, which is the nearest AT of training data in 
$c_1$) to distances $b_1$ and $b_2$ (i.e., distance to $c_2$ measured from 
the reference point), we get a sense of how close to the class boundary the 
new inputs are.
We posit that, for classification problems, inputs that are closer to
class boundaries are more surprising and valuable in terms of test input 
diversity.
On the other hand, for tasks without any boundaries between inputs, such as
prediction of appropriate steering angle for autonomous driving car, DSA may not be 
easily applicable. With no class boundaries, an AT of a new input being far from that of another training input does not guarantee that the new input is 
surprising, as the AT may still be located in crowded parts of the AT space.
Consequently, we only apply DSA for classification tasks, for which it can
be more effective than LSA (see Section~\ref{sec:rq1} and~\ref{sec:rq2} for 
more details).

Let us assume that a DL system $\mathbf{D}$, which consists of a set of 
neurons $\mathbf{N}$, is trained for a classification task with a set of 
classes $C$, using a training dataset $\mathbf{T}$. Given the set of 
activation traces $A_\mathbf{N}(\mathbf{T})$, a new input $x$, and a 
predicted class of the new input $c_x \in C$, we define the reference point 
$x_a$ to be 
the closest neighbour of $x$ that shares the same class. The distance between $x$ and $x_a$ follows from the definition:

\begin{equation}
    \begin{split}
        &x_a = \argmin_{\mathbf{D}(x_i) = c_x}\|\alpha_\mathbf{N}(x) - \alpha_\mathbf{N}(x_i)\|,\\
        &dist_a = \|\alpha_\mathbf{N}(x) - \alpha_\mathbf{N}(x_a)\|
    \end{split}
\label{eq:dist_a}
\end{equation}

Subsequently, from $x_a$, we find the closest neighbour of $x_a$ in a class 
other than $c_x$, $x_b$, and the distance $dist_b$, as follows:

\begin{equation}
    \begin{split}
        &x_b = \argmin_{\mathbf{D}(x_i) \in C \setminus \{c_x\}}\|\alpha_\mathbf{N}(x_a) - \alpha_\mathbf{N}(x_i)\|,\\
        &dist_b = \|\alpha_\mathbf{N}(x_a) - \alpha_\mathbf{N}(x_b)\|
    \end{split}
\label{eq:dist_b}
\end{equation}

Intuitively, DSA aims to compare the distance from the AT of 
a new input $x$ to known ATs belonging to its own class, $c_x$, to the known 
distance between ATs in class $c_x$ and ATs in other classes in $C \setminus \{c_x\}$. 
If the former is relatively larger than the latter, $x$ would be a 
surprising input for class $c_x$ to the classifying DL system $\mathbf{D}$. 
While there are multiple ways to formalise this we select a simple one 
and calculate DSA as the ratio between $dist_a$ and $dist_b$. Investigation
of more complicated formulations is left as future work.

\begin{equation}
    DSA(x) = \frac{dist_a}{dist_b}    
\label{eq:dsa}
\end{equation}

\subsection{Surprise Coverage}
\label{sec:supcov}

Given a \emph{set} of inputs, we can also measure the range of SA values the 
set \emph{covers}, called Surprise Coverage (SC). Since both LSA and DSA are 
defined in continuous spaces, we use bucketing to discretise the space of 
surprise and define both Likelihood-based 
Surprise Coverage (LSC) and Distance-based Surprise Coverage (DSC). Given an 
upper bound of $U$, and buckets $B = \{b_1, b_2, ... , b_n\}$ that divide 
$(0, U]$ into $n$ SA segments, SC for a set of inputs $X$ is defined as 
follows:

\begin{equation}
    SC(X) = \frac{|\{b_i \mid \exists x \in X : SA(x) \in (U\cdot\frac{i-1}{n}, U\cdot\frac{i}{n}]\}|}{n}
\label{eq:D_coverage}
\end{equation}

A set of inputs with high SC is a diverse set of inputs ranging from similar to 
those seen during training (i.e., low SA) to very different from what was seen 
during training (i.e., high SA). We argue that an input set for a DL system 
should not only be diversified, but \emph{systematically} diversified considering SA.
Recent results 
also validate this notion by showing 
that more distant test inputs were more likely to lead to exceptions but might not be as relevant for testing~\cite{Poulding2017aa}.

While we use the term \emph{cover} and \emph{coverage}, the implications of SA 
based coverage is different from the traditional structural 
coverage. First, unlike most of the structural coverage criteria, there is no 
finite set of targets to cover, as in statement or branch coverage: an input 
can, at least in theory, be arbitrarily surprising. 
However, an input with arbitrarily high SA 
value may simply be irrelevant, or at least less interesting, to the problem domain (e.g., an image of a 
traffic sign will be irrelevant to the testing of animal photo classifiers). 
As such, SC can only be measured with respect to pre-defined 
upper bound, in the same way the theoretically infinite path coverage is 
bounded by a parameter~\cite{Zhu1997rt}. Second, SC does not render itself to 
a combinatorial set cover problem, which the test suite minimisation is often
formulated into~\cite{Yoo:2010fk}. This is because a single input yields only 
a single SA value and cannot belong to multiple SA buckets. The sense of 
redundancy with respect to SC as a coverage criteria is weaker than that of 
structural coverage, for which a single input can cover multiple targets. 
While we aim to show that SA can guide the better selection of inputs, rigorous 
study of optimisation of test suites for DL systems remains a future work.
However, as we show 
with our empirical studies, SC can still guide test input selection.

\begin{table*}
    \centering
    \caption{List of datasets and models used in the study.}
    \label{table:dataset}
    
      \begin{tabular}{lp{5cm}p{3cm}lp{2.5cm}p{1.5cm}}
        \toprule
        Dataset                                               & Description                                                                                                                                                                                                                & DNN Model                                                                                                       & \# of Neuron              & Synthetic Inputs                                                                               & Performance                          \\
        \midrule
        MNIST                                                 & Handwritten digit images composed of 50,000 images for training and 10,000 images for test.                                                                                                                                & A five layer ConvNet with max-pooling and dropout layers. & 320                       & FGSM, BIM-A, BIM-B, JSMA, C\&W.                                                       & 99.31\% (Accuracy)                          \\
  
        \midrule
        CIFAR-10                                              & Object recognition dataset in ten different classes composed of 50,000 images for training and 10,000 images for test.                                                                                                      & A 12 layer ConvNet with max-pooling and dropout layers.                                               & 2,208                     & FGSM, BIM-A, BIM-B, JSMA, C\&W.                                                       & 82.27\% (Accuracy)                          \\
  
        \midrule
        \multirow{2}{2cm}{Udacity Self-driving Car Challenge} & \multirow{2}{5cm}{Self-driving car dataset that contains camera images from the vehicle, composed of 101,396 images for training and 5,614 images for test. The goal of the challenge is to predict steering wheel angle.} & Dave-2~\cite{Bojarski2016ak} architecture from Nvidia.                               & 1,560 & DeepXplore's test input generation via joint optimization. & 0.09 (MSE) \\\cmidrule{3-6}
                                                              &                                                                                                                                                                                                                            & Chauffeur~\cite{Chauffeur} architecture with CNN and LSTM.         & 1,940 & DeepTest's combined transformation.                       & 0.10 (MSE) \\
        \bottomrule
      \end{tabular}
\end{table*}

\section{Research Questions}
\label{sec:rqs}

Our empirical evaluation is designed to answer the following research 
questions.

\vspace{0.5em}

\noindent\textbf{RQ1. Surprise:} Is \name capable of capturing the 
relative surprise of an input of a DL system?

\vspace{0.5em}

We provide answers to RQ1 from different angles. First, we compute the SA of
each test input included in the original dataset, and see if a DL classifier 
finds inputs with higher surprise more difficult to correctly classify. We
expect more surprising input to be harder to correctly classify. Second, we
evaluate whether it is possible to detect adversarial examples based on SA 
values, as we expect adversarial examples to be more surprising as well as to
cause different behaviours of DL systems. Using different techniques, multiple 
sets of adversarial examples are generated and compared by their SA values. 
Finally, we train adversarial example classifiers using logistic
regression on SA values. For each adversarial attack strategy, we
generate 10,000 adversarial examples using 10,000 original test images provided
by MNIST and CIFAR-10. Using 1,000 original test images and 1,000 adversarial
examples, all chosen randomly, we train the logistic regression classifiers.
Finally, we evaluate the trained classifiers using the remaining 9,000 original
test images and 9,000 adversarial examples. 
If SA values correctly capture the 
behaviour of DL systems, we expect the SA based classifiers to successfully 
detect adversarial examples. We use Area Under Curve of Receiver Operator 
Characteristics (ROC-AUC) for evaluation as it captures both
true and false positive rates~\cite{carlini2017adversarial}.

\vspace{0.5em}

\noindent\textbf{RQ2. Layer Sensitivity:} Does the selection of layers of 
neurons used for SA computation have any impact on how accurately SA reflects
the behaviour of DL systems?

\vspace{0.5em}

Bengio et al. suggest that deeper layers represent higher level features of 
the input~\cite{Bengio2012aa}: subsequent work that introduced KDE based 
adversarial example detection technique~\cite{Feinman2017aa} assumes 
the deepest (i.e., the last hidden) layer to contain the most information 
helpful for detection. We evaluate this assumption in the 
context of SA by calculating LSA and DSA of all individual layers, and 
subsequently by comparing adversarial example classifiers trained on SA from 
each layer.

\vspace{0.5em}

\noindent\textbf{RQ3. Correlation:} Is SC correlated to existing coverage criteria for DL systems?
\vspace{0.5em}

In addition to capturing input surprise, we want SC to be consistent with
existing coverage criteria based on counting aggregation. If not, there is a 
risk that SC is in fact measuring something other than input diversity.
For this, we check whether SC is correlated with other criteria.
We control the input diversity 
by cumulatively adding inputs generated by different method (i.e., different 
adversarial example generation techniques or input synthesis techniques),
execute the studied DL systems with these input, 
and compare the observed changes of various coverage criteria including
SC and four existing ones: DeepXplore's Neuron Coverage 
(NC)~\cite{Tian2018zn} and three Neuron-level Coverages (NLCs) introduced by 
DeepGauge~\cite{Ma2018ny}: $k$-Multisection Neuron Coverage (KMNC),
Neuron Boundary Coverage (NBC), and Strong Neuron Activation Coverage (SNAC).

For MNIST and 
CIFAR-10, we start from the original test data provided by the dataset 
(10,000 images), and add 1,000 adversarial examples, generated by FGSM, BIM-A, 
BIM-B, JSMA, and C\&W, at each step.
For Dave-2, we start from the original test data (5,614 images) and add 700 
synthetic images generated by DeepXplore at each step. For Chauffeur, each 
step adds 1,000 synthetic images (Set1 to Set3), each produced by 
applying random number of DeepTest transformations.

\vspace{0.5em}

\noindent\textbf{RQ4. Guidance:} Can SA guide retraining of DL 
systems to improve their accuracy against adversarial examples and synthetic 
test inputs generated by DeepXplore?
\vspace{0.5em}

To evaluate whether \name can guide additional training of existing DL systems
with the aim of improved accuracy against adversarial examples, we ask whether
SA can guide the selection of input for additional training.
From the adversarial 
examples and synthesised inputs for these models\footnote{We could not resume 
training of Chauffeur model for additional five epochs, which is why it is 
absent from RQ4.}, we choose four sets of 100 images from four different SA 
ranges. Given $U$ as the upper bound used in RQ3 to compute the SC, we divide
the range of SA $[0, U]$ into four overlapping subsets: the first subset
including the low 25\% SA values ($[0, \frac{U}{4}]$), the second including 
the lower half ($[0, \frac{2U}{4}]$), the third including the lower 75\% ($[0, 
\frac{3U}{4}]$), and finally the entire range, $[0, U]$. These four subsets
are expected to represent increasingly more diverse sets of inputs.
We set the range $R$ to one of these four, randomly sample 100 images from 
each $R$, and train existing models for five additional epochs. Finally,
we measure each model's performance (accuracy for MNIST and CIFAR-10, MSE for 
Dave-2) against the entire adversarial and synthetic inputs, respectively.
We expect retraining with more diverse subset will result in higher 
performance.

\section{Experimental Setup}
\label{sec:setup}
We evaluate \name on four different DL systems using (a) the original test sets, 
(b) adversarial examples generated by five attack strategies, and (c) synthetic 
inputs generated by DeepXplore~\cite{Pei2017qy} and DeepTest~\cite{Tian2018zn}. 
This section describes the studied DL systems and the input generation methods.

\subsection{Datasets and DL Systems}
\label{sec:dataset}

Table~\ref{table:dataset} lists the subject datasets and models of DL systems.
MNIST~\cite{LeCun2010vg} and CIFAR-10~\cite{Krizhevsky2014nu} are widely used 
datasets for machine learning research, each of which is a collection of 
images in ten different classes.
For MNIST, we adopt the widely studied five layer Convolutional Neural Network 
(ConvNet) with max-pooling and dropout layers
and train it to achieve 99.31\% accuracy on the provided test set. Similarly, 
the adopted model for CIFAR is a 12-layer ConvNet
with max-pooling and dropout layers, trained to achieve 82.27\% accuracy on the
provided test set.

For evaluation of \name for DL systems in safety critical domains, we use the 
Udacity self-driving car challenge dataset~\cite{Udacitydataset},
which contains a collection of camera images from the driving car. As its aim 
is to predict steering wheel angle,
the model accuracy is measured using Mean Squared Error (MSE) between actual 
and predicted steering angles. We use a pre-trained Dave-2
model~\cite{Bojarski2016ak}, which is a public artefact provided
by DeepXplore\footnote{DeepXplore is available from: \url{https://github.com/peikexin9/deepxplore}.},
and a pre-trained Chauffeur model~\cite{Chauffeur},
made publicly available by the Udacity self-driving car challenge. 
Dave-2 consists of nine layers including five convolutional layers, and 
achieves 0.09 in MSE. 
Chauffeur consists of both a ConvNet and an LSTM sub-model, and achieves
0.10 in MSE.

\subsection{Adversarial Examples and Synthetic Inputs}
\label{sec:inputs}

\name is evaluated using both adversarial examples and synthetic test inputs. 
Adversarial examples are crafted by applying, to the original input, small 
perturbations imperceptible to humans, until the DL system under investigation 
behaves incorrectly~\cite{Goodfellow43405}. We rely on adversarial attacks
to generate input images for MNIST and CIFAR-10: these generated images are
more likely to reveal robustness issues in the DL systems than the test inputs
provided by the original datasets. We use five widely studied attack strategies
to evaluate \name: Fast Gradient Sign Method (FGSM)~\cite{Goodfellow43405}, 
Basic Iterative Method (BIM-A, BIM-B)~\cite{Kura1607},
Jacobian-based Saliency Map Attack (JSMA)~\cite{PapernotMJFCS15},
and Carlini\&Wagner (C\&W)~\cite{CarliniW16a}. Our implementation of these
strategies is based on \texttt{cleverhans}~\cite{papernot2018cleverhans} and
a framework of Ma et al.~\cite{ma2018characterizing}.

For Dave-2 and Chauffeur, we use the state-of-the-art synthetic input
generation algorithms, DeepXplore~\cite{Pei2017qy} and 
DeepTest\cite{Tian2018zn}. Both algorithms are designed to synthesise new
test input from existing ones with the aim of detecting erroneous behaviours in
autonomous driving vehicle.
For Dave-2, we use DeepXplore's input generation via joint optimization 
algorithm, whose aim is to generate inputs that lead multiple DL systems 
trained independently, but using the same training data, to disagree with each 
other. Using Dave-2 and its two variants, Dave-dropout and Dave-norminit,
we collect synthetic inputs generated by lighting effect (Light), occlusion by 
a single black rectangle (SingleOcc), and occlusion by multiple black rectangles 
(MultiOcc). 
For Chauffeur, we synthesise new inputs by iteratively applying 
random transformations provided by DeepTest to original input images: 
translation, scale, shear, rotation, contrast, brightness, and blur.\footnote{At
the time of our experiments, the publicly available version of 
DeepTest did not internally support realistic image transformations such as 
fog and rain effects.}

\begin{table}[ht]
    \centering
    \caption{Configurations for RQ3.}
    \label{table:configurations}
    \scalebox{0.85}{
    \begin{tabular}{l|r|r|lrr|rr}
    \toprule
    DNN      & NC & NLCs & \multicolumn{3}{c|}{LSC}   & \multicolumn{2}{c}{DSC}             \\
    Model   & $th$ & $k$ & layer & $n$ & $ub$ & $n$ & $ub$ \\
    \midrule
    MNIST     & 0.5 & 1,000         & activation\_3 & 1,000 & 2,000 & 1,000 & 2.0 \\
    CIFAR-10  & 0.5 & 1,000         & activation\_3 & 1,000 & 100   & 1,000 & 2.0 \\
    Dave-2    & 0.5 & 1,000         & block1\_conv2 & 1,000 & 150 & \multicolumn{2}{c}{N/A}           \\
    Chauffeur & 0.5 & 1,000         & convolution2d\_11 & 1,000 & 5 & \multicolumn{2}{c}{N/A}           \\
    \bottomrule
    \end{tabular}
    }
\end{table}

\subsection{Configurations}
\label{sec:configurations}

For all research questions, the default activation variance threshold for LSA 
is set to $10^{-5}$, and the bandwidth for KDE is set using Scott's 
Rule~\cite{scott2015multivariate}. The remaining of this Section details 
RQ specific configurations. For RQ1, we use the activation\_2 layer for MNIST, 
and activation\_6 for CIFAR-10, when computing LSA values. Computation of LSA 
based on all neurons is computationally infeasible due to precision loss.
For RQ2, we set the activation variance threshold for layers activation\_7
and activation\_8 of CIFAR-10 to $10^{-4}$, which reduces the number of neurons
used for the computation of LSA and, consequently, the computational cost.
For computation of other coverage criteria in RQ3, we use the configurations
in Table~\ref{table:configurations}. The threshold of NC is set to 0.5. 
For NLCs, we all set the number of sections ($k$) to 1,000.
For LSC and DSC, we manually choose the layer, the number of buckets ($n$),
and the upper bound ($ub$).
For RQ4, the layers chosen for MNIST and CIFAR-10 are activation\_3 and activation\_5 
respectively. We perform 20 runs of retraining for each subject and report the
statistics.

All experiments were performed on machines equipped with Intel i7-8700 CPU, 32GB
RAM, running Ubuntu 16.04.4 LTS. MNIST and CIFAR-10 are implemented using
Keras v.2.2.0.

\begin{figure}[!ht]
    \centering
    \begin{minipage}[t]{.43\textwidth}
        \centering
        \includegraphics[width=\textwidth]{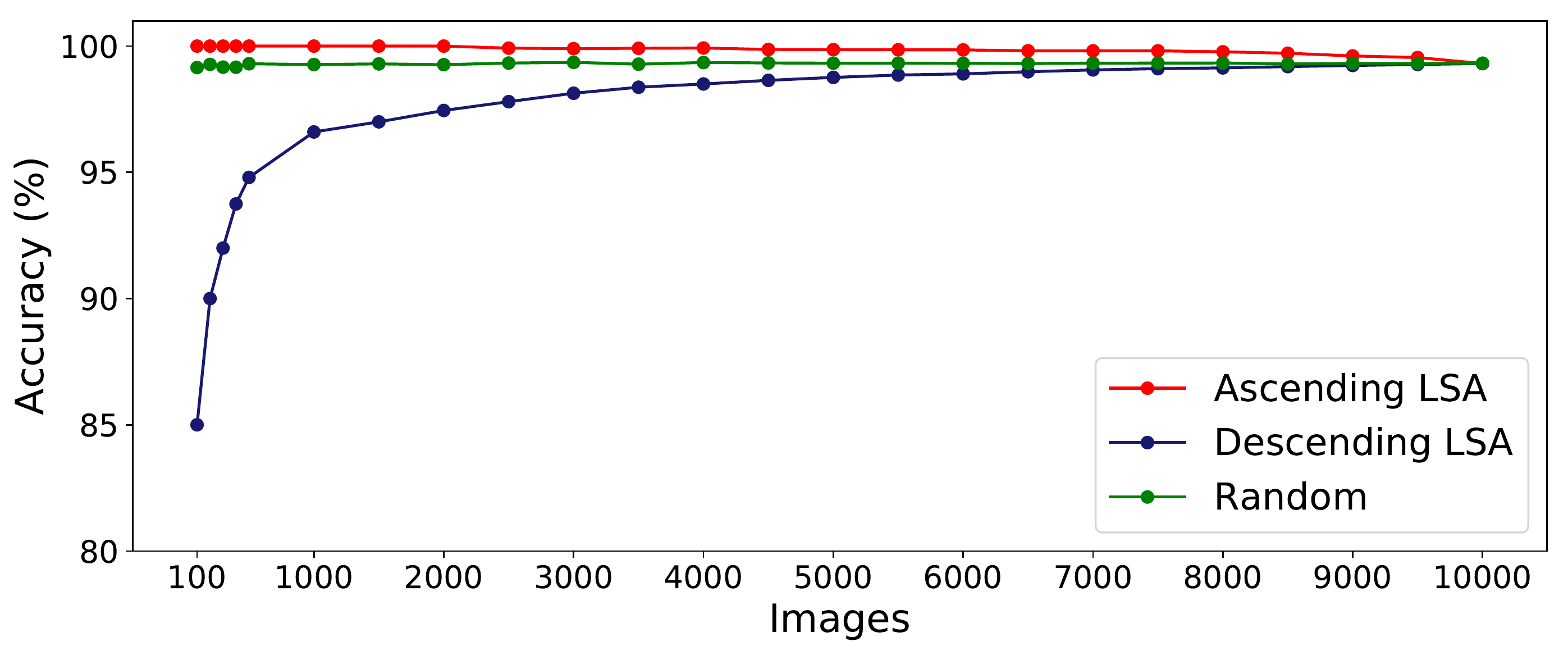}
        \subcaption{Selected test inputs based on LSA in MNIST}\label{fig:RQ1_test_acc_lsa_mnist}    
        \vspace{0.5em}
    \end{minipage}        
    \begin{minipage}[t]{.43\textwidth}
        \centering
        \includegraphics[width=\textwidth]{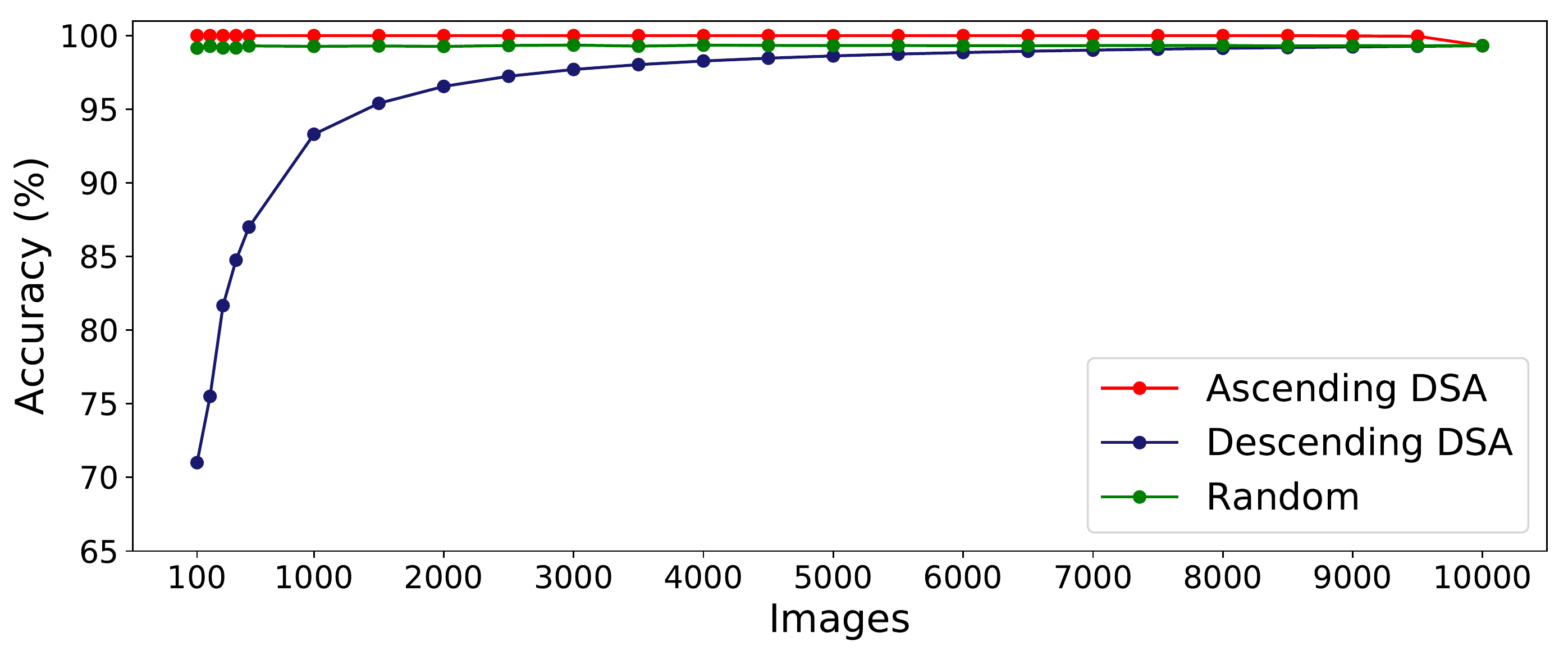}        
        \subcaption{Selected test inputs based on DSA in MNIST}\label{fig:RQ1_test_acc_dsa_mnist}
        \vspace{0.5em}
    \end{minipage}           
    \begin{minipage}[t]{.43\textwidth}
        \centering
        \includegraphics[width=\textwidth]{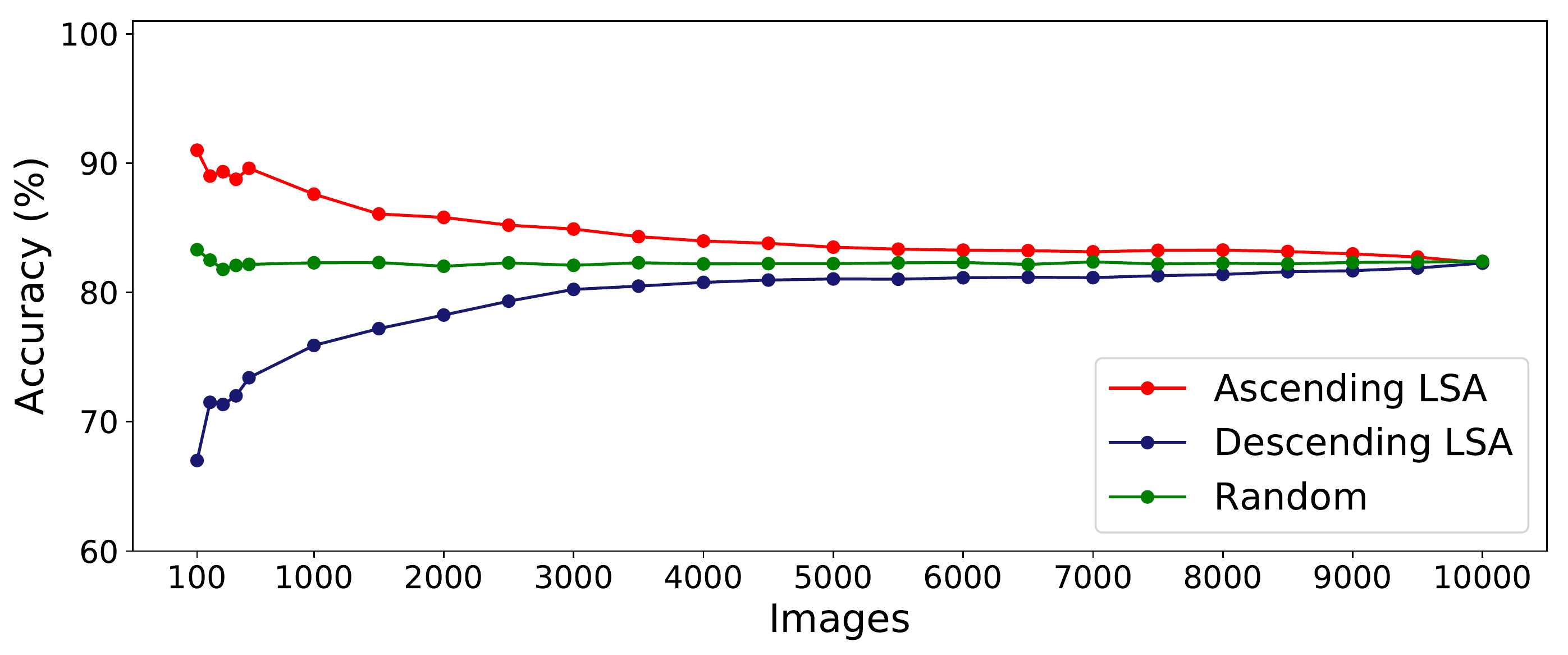}
        \subcaption{Selected test inputs based on LSA in CIFAR-10}\label{fig:RQ1_test_acc_lsa_cifar}
        \vspace{0.5em}
    \end{minipage}           
    \begin{minipage}[t]{.43\textwidth}
        \centering
        \includegraphics[width=\textwidth]{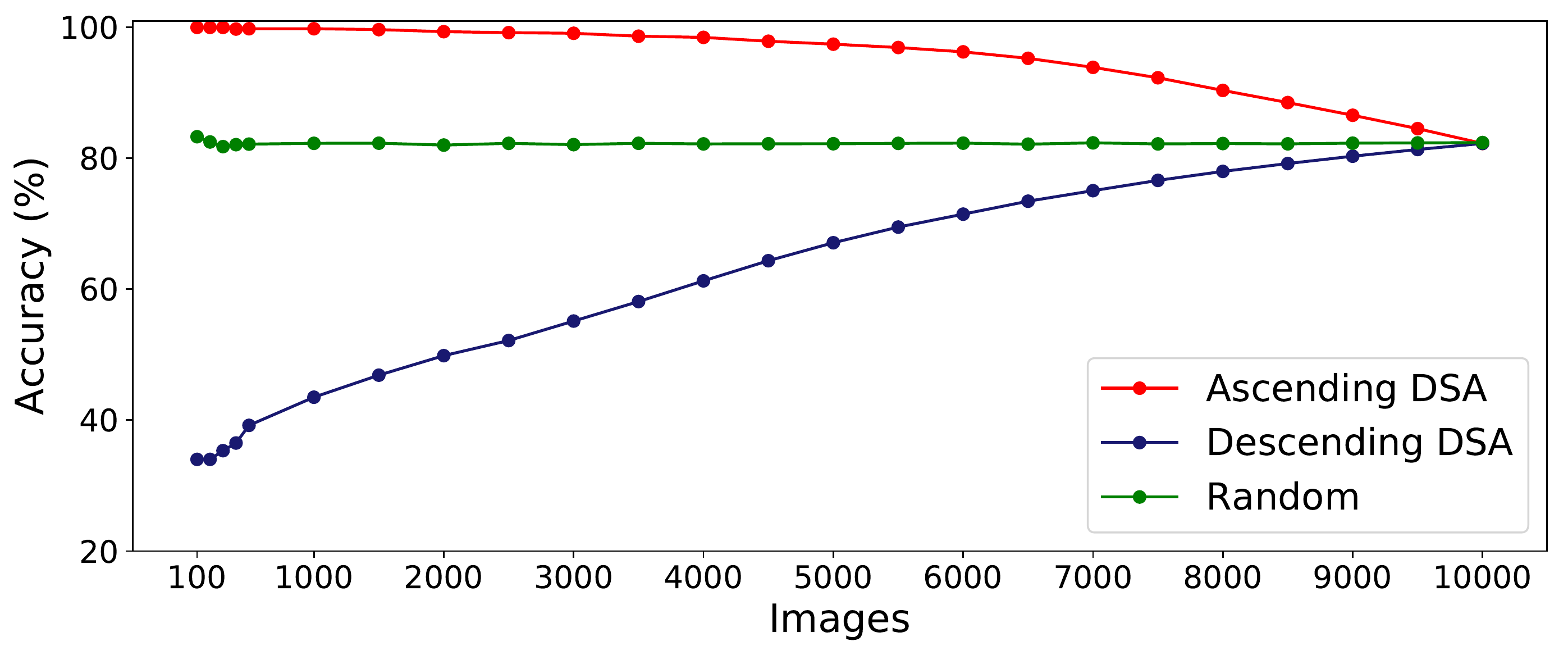}
        \subcaption{Selected test inputs based on DSA in CIFAR-10}\label{fig:RQ1_test_acc_dsa_cifar}        
    \end{minipage}   
    \vspace{0.5em}
    \caption{Accuracy of test inputs in MNIST and CIFAR-10 dataset, selected from
    the input with the lowest SA, increasingly including inputs with higher SA,
    and vice versa (i.e., from the input with the highest SA to inputs with lower SA).
    \label{fig:test_accuracy}}
\end{figure}

\section{Result}
\label{sec:result}


\subsection{Input Surprise (RQ1)}
\label{sec:rq1}

\begin{figure}[!htb]
    \centering
    \begin{minipage}{0.45\textwidth}
        \centering
        \minipage{0.32\textwidth}
        \includegraphics[width=\linewidth]{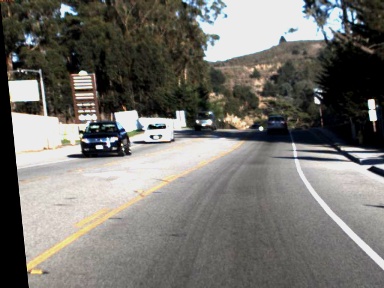}        
        \endminipage\hfill
        \minipage{0.32\textwidth}
        \includegraphics[width=\linewidth]{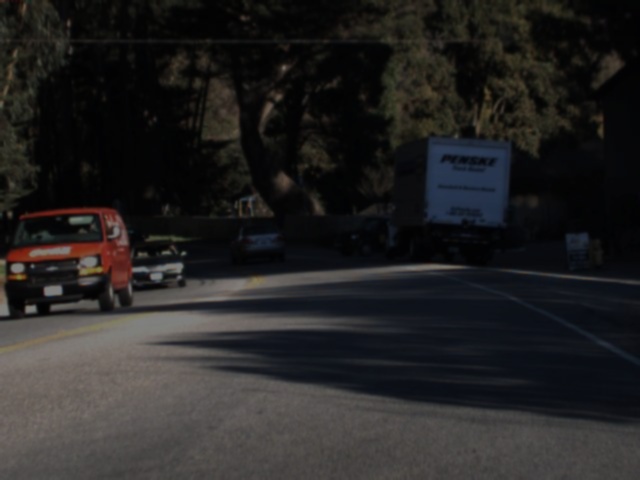}        
        \endminipage\hfill
        \minipage{0.32\textwidth}%
        \includegraphics[width=\linewidth]{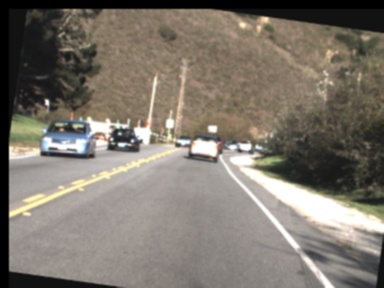}        
        \endminipage
        \subcaption{Low LSA}
        \vspace{0.5em}
    \end{minipage}
    \begin{minipage}{0.45\textwidth}
        \minipage{0.32\textwidth}
        \includegraphics[width=\linewidth]{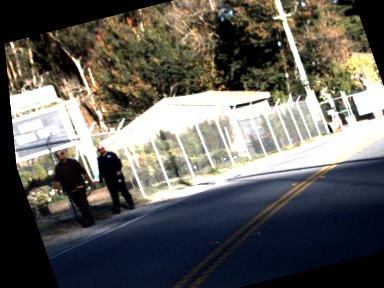}        
        \endminipage\hfill
        \minipage{0.32\textwidth}
        \includegraphics[width=\linewidth]{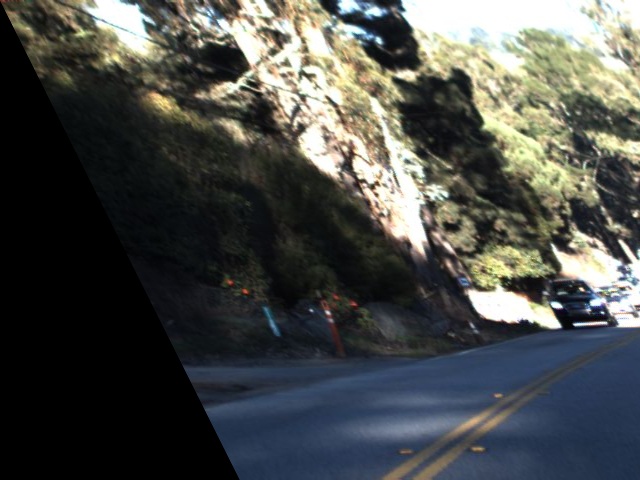}        
        \endminipage\hfill
        \minipage{0.32\textwidth}%
        \includegraphics[width=\linewidth]{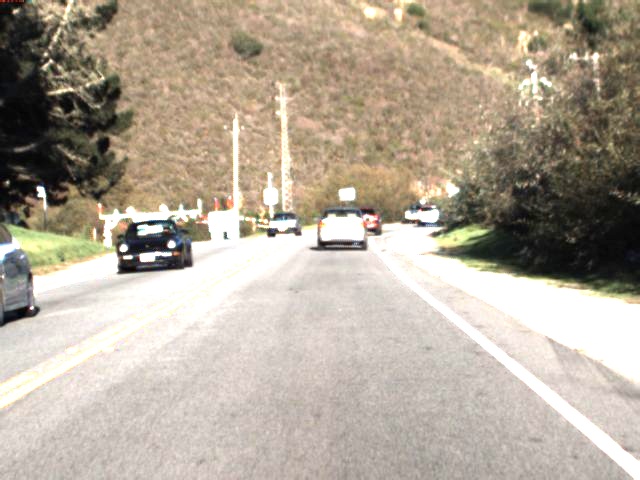}        
        \endminipage
        \subcaption{Medium LSA}
        \vspace{0.5em}
    \end{minipage}
    \begin{minipage}{0.45\textwidth}
        \minipage{0.32\textwidth}
        \includegraphics[width=\linewidth]{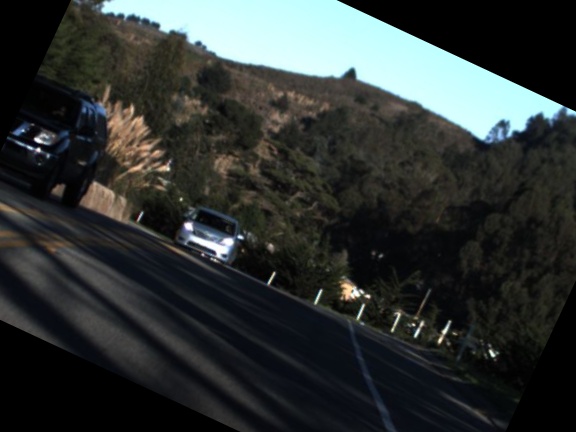}        
        \endminipage\hfill
        \minipage{0.32\textwidth}
        \includegraphics[width=\linewidth]{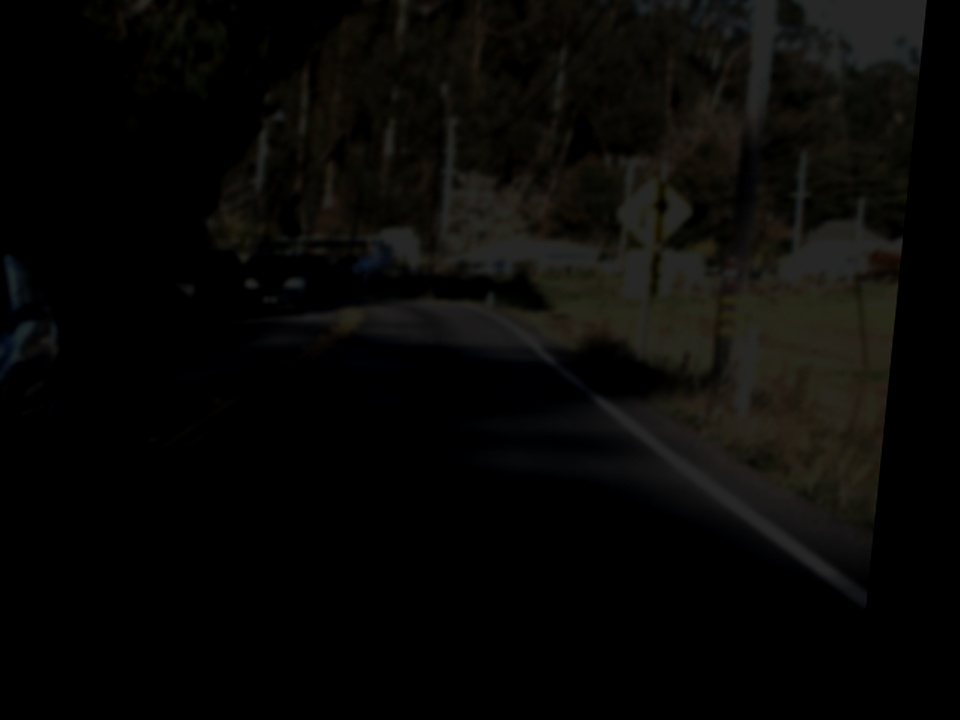}        
        \endminipage\hfill
        \minipage{0.32\textwidth}%
        \includegraphics[width=\linewidth]{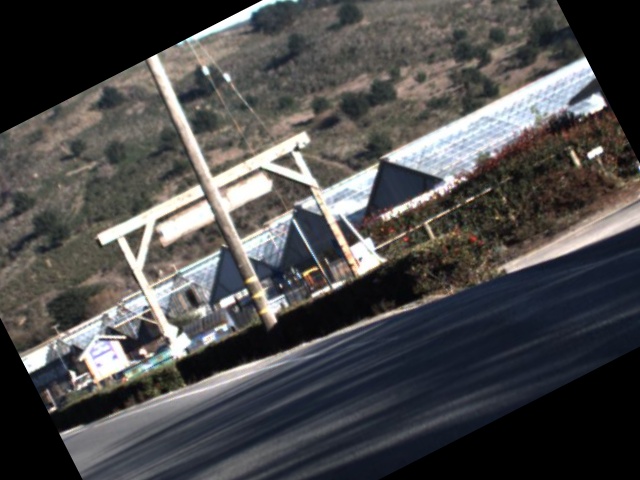}        
        \endminipage
        \subcaption{High LSA}        
    \end{minipage}
    \caption{Synthetic images for Chauffeur model generated by DeepTest. 
    Images with higher LSA values tend to be harder to recognise and interpret visually.
    \label{fig:images_by_surp}}    
\end{figure}

Figure~\ref{fig:test_accuracy} shows how the classification accuracy
changes when we classify sets of images of growing sizes from the test inputs included 
in the MNIST and CIFAR-10 dataset. The sets of images corresponding to the red dots (Ascending SA) start 
with images with the lowest SA, and increasingly include images with higher SA in 
the ascending order of SA; the sets of images corresponding to the blue dots 
grow in the opposite direction (i.e., from images with the highest SA to lower 
SA). As a reference, the green dots show the mean accuracy of randomly growing sets across 
20 repetitions. It is clear that including images with higher LSA values, i.e., more surprising images,
leads to lower accuracy. For visual confirmation on another dataset, we also chose sets of 
inputs synthesised for Chauffeur by DeepTest, from three distinct levels of 
LSA values: Figure~\ref{fig:images_by_surp} shows that the higher the LSA values 
are, the harder it is to recognise images visually. 
Both quantitatively and visually, the observed trend supports 
our claim that \name captures input surprise: even for unseen inputs, SA can 
measure how surprising the given input is, which is directly related to the 
performance of the DL system.

\begin{figure}[ht]
    \centering
        \includegraphics[width=0.75\columnwidth]{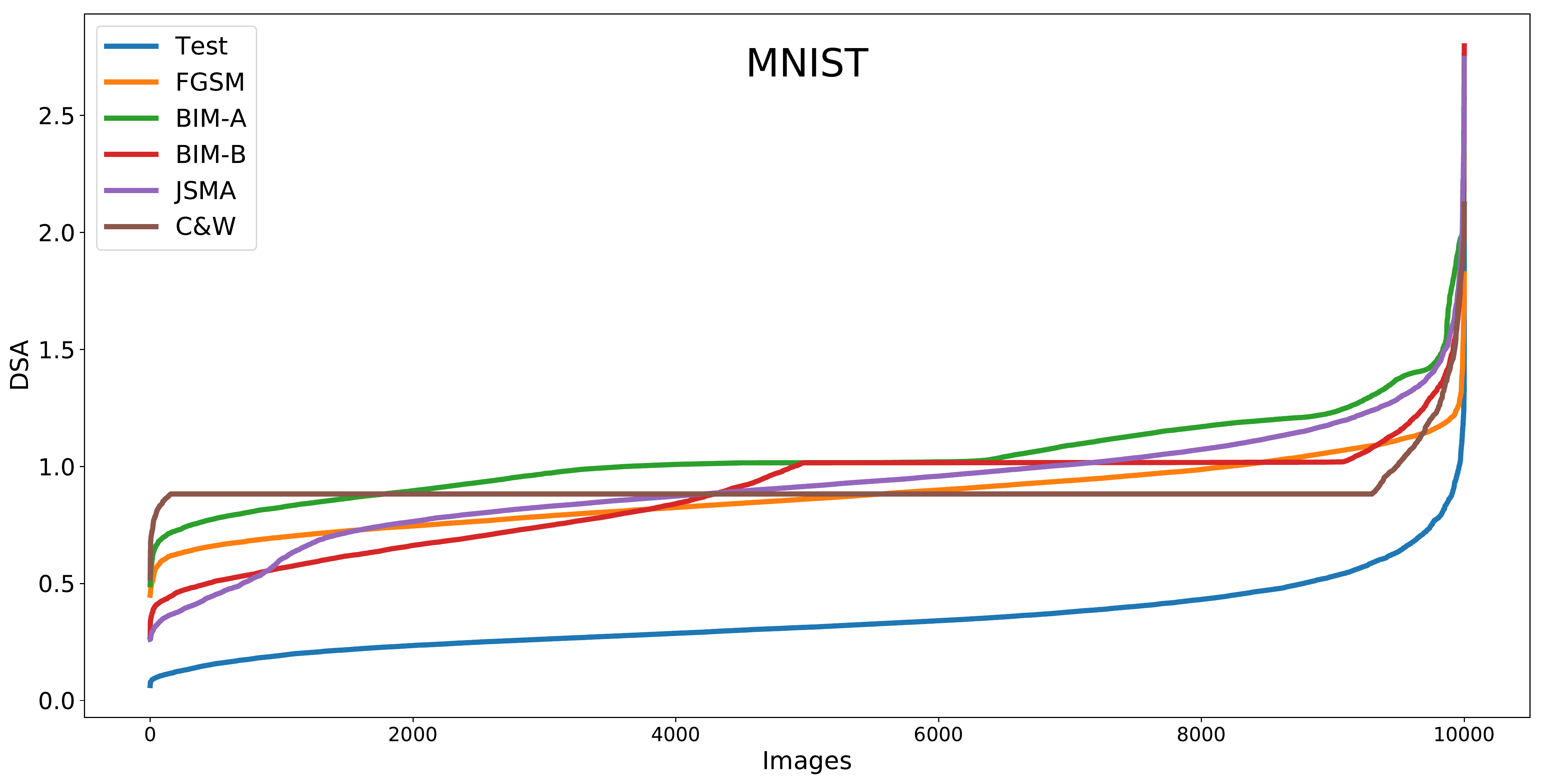}
        \includegraphics[width=0.75\columnwidth]{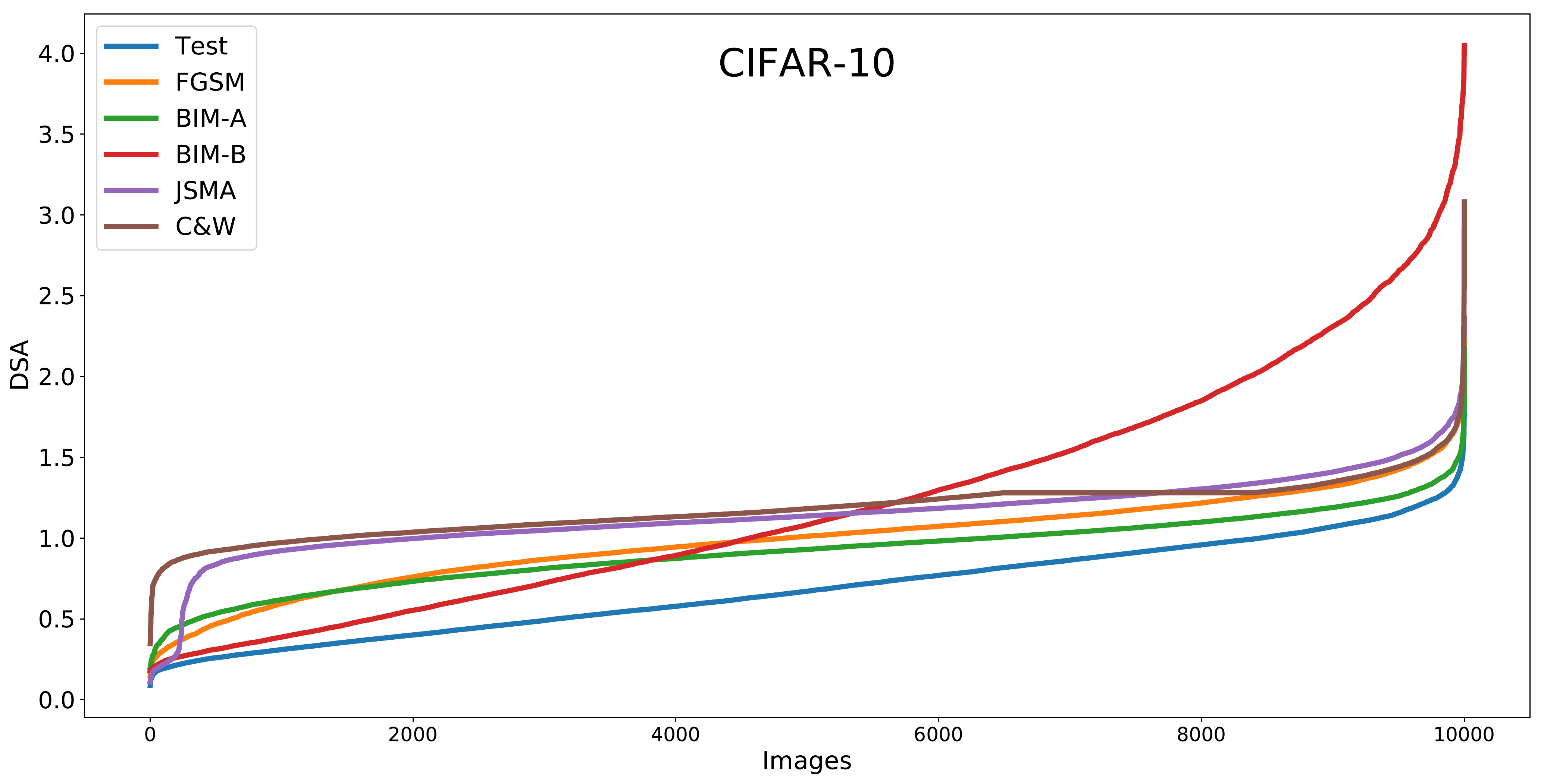}
    \caption{Sorted DSA values of adversarial examples for MNIST and CIFAR-10.\label{fig:RQ1_distance}}
\end{figure}

\begin{figure*}[ht]
    \centering
    \begin{minipage}[t]{.85\textwidth}
        \centering
        \includegraphics[width=\textwidth]{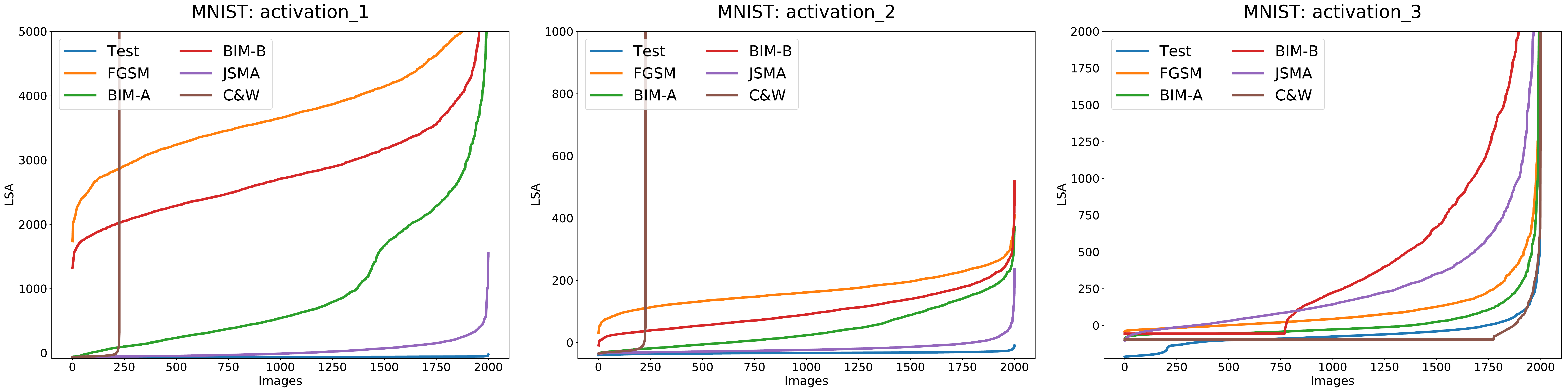}
        \vspace{0.1em}
    \end{minipage}       
    \hfill
    \begin{minipage}[t]{.85\textwidth}
        \centering
        \includegraphics[width=\textwidth]{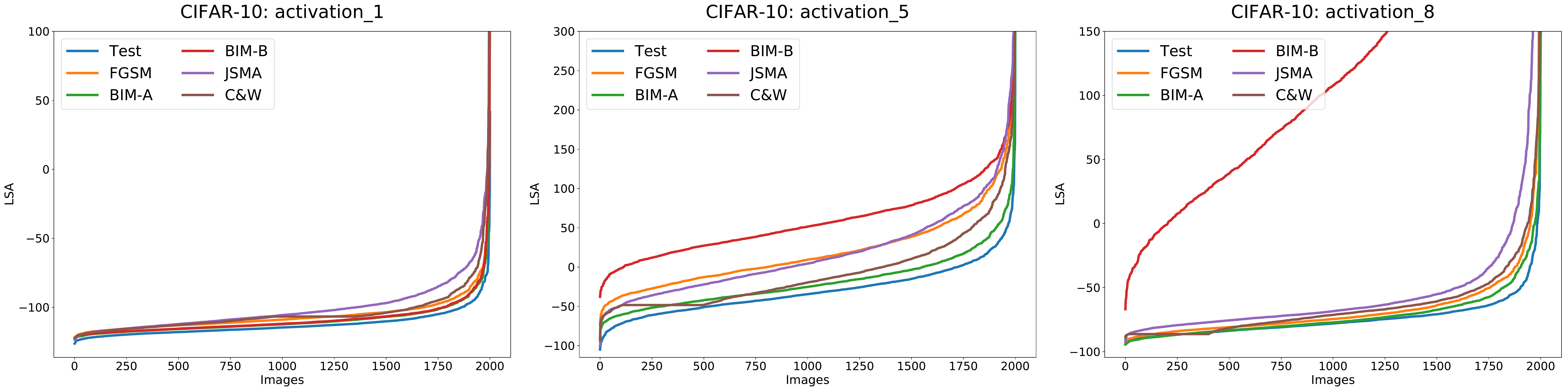}
    \end{minipage} 
    \caption{Sorted LSA of randomly selected 2,000 adversarial examples for MNIST and CIFAR-10 from different layers}
    \label{fig:RQ1_layer_selection}
\end{figure*}

Figure~\ref{fig:RQ1_distance} shows plots of sorted DSA values of 10,000
adversarial examples, generated by each of the five techniques, as well 
as the original test inputs. 
Figure~\ref{fig:RQ1_layer_selection} contains similar plots based on LSA 
values of 2,000 randomly selected adversarial examples and the original test set,
from different layers of MNIST and CIFAR-10.
For both MNIST and CIFAR-10, the test inputs provided with the datasets
(represented in blue colour) tend to be the least surprising, whereas the 
majority of adversarial examples are clearly separated from the test inputs by 
their higher SA values. This supports our claim that \name can 
capture the differences in DL system's behaviours for adversarial 
examples.

Finally, Table~\ref{table:rq1_dsa_auc} shows the ROC-AUC results of DSA-based 
classification using all neurons in MNIST and CIFAR-10.\footnote{LSA-based 
classification is only possible for subsets of neurons due to the 
computational cost of KDE; hence we introduce the results of LSA-based 
classification when answering the impact of layer selection for RQ2.} 
The results show that the gap in DSA values observed in 
Figure~\ref{fig:RQ1_distance} can be used to classify adversarial examples 
with high accuracy. For the relatively simpler MNIST model, the DSA-based classifier
can detect adversarial examples with ROC-AUC ranging from 96.97\% to 99.38\%.
The DSA-based classification for the more complicated CIFAR-10 model shows
lower ROC-AUC values, but answers to RQ2 suggest that DSA from specific
layers can produce significantly higher accuracy (see Section~\ref{sec:rq2}).

\begin{table}[ht]
    \centering
    \caption{ROC-AUC of DSA-based classification of adversarial examples for MNIST and CIFAR-10\label{table:rq1_dsa_auc}}
    \begin{tabular}{llllll}
      \toprule
      Dataset  & FGSM    & BIM-A   & BIM-B   & JSMA    & C\&W    \\
      \midrule      
      MNIST    & 98.34\% & 99.38\% & 96.97\% & 97.10\% & 99.04\% \\      
      CIFAR-10 & 76.81\% & 72.93\% & 71.66\% & 88.96\% & 92.84\% \\
      \bottomrule      
    \end{tabular}    
\end{table}

Based on three different analyses, the answer to RQ1 is that \textbf{\name can capture the
relative surprise of inputs}. Inputs with higher SA are harder to correctly
classify; adversarial examples show higher SA values and can be classified
based on SA accordingly.

\subsection{Impact of Layer Selection (RQ2)}
\label{sec:rq2}

Table~\ref{table:rq2_sa_roc_auc_per_layer_mnist} shows the ROC-AUC of 
classification of adversarial examples, resulting in each row corresponding to a 
classifier trained on LSA and DSA from a specific layer in MNIST, 
respectively. Rows are ordered by their
depth, i.e., activation\_3 is the deepest and the last hidden layer in MNIST. 
The highest ROC-AUC values for each attack strategy are typeset in bold.
For MNIST, there is no clear evidence that the deepest layer is the most
effective.

\begin{table}[ht]
    \centering
    \caption{ROC-AUC results of SA per layers on MNIST.}
    \label{table:rq2_sa_roc_auc_per_layer_mnist}
\scalebox{0.90}{
    \begin{tabular}{cllllll}
    \toprule
    SA & Layer        & FGSM    & BIM-A   & BIM-B   & JSMA   & C\&W    \\
    \midrule    
    \multirow{4}{*}{LSA}&activation\_1 & \textbf{100.00\%} & \textbf{99.94\%}  & \textbf{100.00\%} & 98.17\% & 99.48\% \\
    &activation\_2 & \textbf{100.00\%} & 99.46\%  & \textbf{100.00\%} & 94.42\% & 99.23\% \\
    &pool\_1        & \textbf{100.00\%} & 99.73\%  & \textbf{100.00\%} & \textbf{99.08\%} & \textbf{99.61\%} \\
    &activation\_3 & 93.29\%  & 81.70\%  & 86.73\%  & 94.45\% & 37.96\% \\
    \midrule            
    \multirow{4}{*}{DSA} &activation\_1 & \textbf{100.00\%} & \textbf{99.85\%}  & \textbf{100.00\%} & 97.79\% & 99.39\% \\
    &activation\_2 & \textbf{100.00\%} & 99.39\%  & 99.99\%  & 97.59\% & \textbf{99.69\%} \\
    &pool\_1        & \textbf{100.00\%} & 99.32\%  & 99.99\%  & \textbf{98.21\%} & \textbf{99.69\%} \\
    &activation\_3 & 98.45\%  & 99.43\%  & 97.40\%  & 97.07\% & 99.10\% \\
    \bottomrule
    \end{tabular}
}

\end{table}

The cases for which ROC-AUC is 100\% can be explained by 
Figure~\ref{fig:RQ1_layer_selection}: LSA values from activation\_1 of MNIST,
for example, show a clear separation between the original test inputs and 
FGSM, BIM-A, or BIM-B: by choosing an appropriate threshold, it is possible to
completely separate test inputs from adversarial examples. Similarly, the plot of LSA
from activation\_3 of MNIST shows that C\&W LSA line crossing with that of 
the original test data (i.e., C\&W adversarial examples are less surprising
than the original test data): this results in the low ROC-AUC value of 37.96\%.

Table~\ref{table:rq2_sa_roc_auc_per_layer_cifar} contains the ROC-AUC values
of LSA- and DSA-based classifiers, trained on each layer of the CIFAR-10 model:
for each attack strategy, the highest ROC-AUC values are typeset in bold.
Interestingly, LSA and DSA show different trends with CIFAR-10. With LSA, there
is no strong evidence that the deepest layer produces the most accurate 
classifiers. However, with DSA, 
the deepest layer produces the most accurate classifiers for three out of 
five attack strategies (BIM-B, JSMA, and C\&W), while the second deepest layer 
produces the most accurate classifier for BIM-A. More importantly, per-layer
DSA values produce much more accurate classification results than all neuron
DSA values, as can be seen in the comparison between Table~\ref{table:rq1_dsa_auc}
and Table~\ref{table:rq2_sa_roc_auc_per_layer_mnist} \& 
\ref{table:rq2_sa_roc_auc_per_layer_cifar}.
Identical models have been used to produce results in Tables above.

\begin{table}[ht]
\centering
\caption{ROC-AUC results of SA per layers on CIFAR-10.\label{table:rq2_sa_roc_auc_per_layer_cifar}}
\scalebox{0.92}{
    \begin{tabular}{cllllll}    
        \toprule
        SA & Layer        & FGSM   & BIM-A   & BIM-B   & JSMA   & C\&W    \\
        \midrule
        \multirow{11}{*}{LSA} &activation\_1 & 72.91\% & 61.59\%  & 63.30\%  & 76.85\% & 74.01\% \\
        &activation\_2 & 89.59\% & 62.17\%  & 73.20\%  & 80.33\% & 75.98\% \\
        &pool\_1        & \textbf{93.31\%} & 61.79\%  & 78.89\%  & 82.64\% & 73.48\% \\
        &activation\_3 & 86.75\% & \textbf{62.69\%}  & 76.93\%  & 80.33\% & 79.02\% \\
        &activation\_4 & 83.31\% & 62.73\%  & 86.15\%  & 80.86\% & \textbf{80.42\%} \\
        &pool\_2        & 82.82\% & 61.16\%  & 89.69\%  & 80.61\% & 73.85\% \\
        &activation\_5 & 83.80\% & 60.64\%  & 96.31\%  & 79.56\% & 64.60\% \\
        &activation\_6 & 63.85\% & 51.90\%  & 99.74\%  & 66.99\% & 60.40\% \\
        &pool\_3        & 63.46\% & 51.86\%  & \textbf{99.77\%}  & 67.62\% & 56.21\% \\
        &activation\_7 & 67.96\% & 61.09\%  & 92.18\%  & \textbf{83.02\%} & 76.85\% \\
        &activation\_8 & 59.28\% & 52.66\%  & 99.60\%  & 73.26\% & 62.15\% \\
        \midrule        
        \multirow{11}{*}{DSA}&activation\_1 & 65.00\% & 62.25\%  & 61.57\%  & 73.85\% & 79.09\% \\
        &activation\_2 & 77.63\% & 64.73\%  & 67.95\%  & 78.16\% & 81.59\% \\
        &pool\_1        & 80.22\% & 64.89\%  & 70.94\%  & 78.96\% & 82.03\% \\
        &activation\_3 & \textbf{83.25\%} & 68.48\%  & 73.49\%  & 79.89\% & 84.16\% \\
        &activation\_4 & 81.77\% & 68.94\%  & 77.94\%  & 80.55\% & 84.62\% \\
        &pool\_2        & 82.51\% & 69.28\%  & 81.43\%  & 80.92\% & 84.81\% \\
        &activation\_5 & 81.45\% & 70.29\%  & 83.28\%  & 82.15\% & 85.15\% \\
        &activation\_6 & 71.71\% & 70.92\%  & 71.15\%  & 84.05\% & 85.42\% \\
        &pool\_3        & 71.75\% & 70.35\%  & 74.65\%  & 83.57\% & 85.17\% \\
        &activation\_7 & 71.04\% & \textbf{71.44\%}  & 81.46\%  & 89.94\% & 92.98\% \\
        &activation\_8 & 70.35\% & 70.65\%  & \textbf{90.47\%}  & \textbf{90.46\%} & \textbf{94.53\%} \\
        \bottomrule
    \end{tabular}
}
\end{table}

Based on these results, we answer RQ2 that \textbf{DSA is sensitive to the
selection of layers it is computed from, and benefits from choosing the deeper
layer}. However, for LSA, there is no clear evidence supporting the deeper layer
assumption. The layer sensitivity varies across different adversarial example
generation strategies.

\subsection{Correlation between SC and Other Criteria (RQ3)}
\label{sec:rq3}

\begin{figure}[ht]
    \centering
    \begin{minipage}[t]{.35\textwidth}
        \centering
        \includegraphics[width=\textwidth]{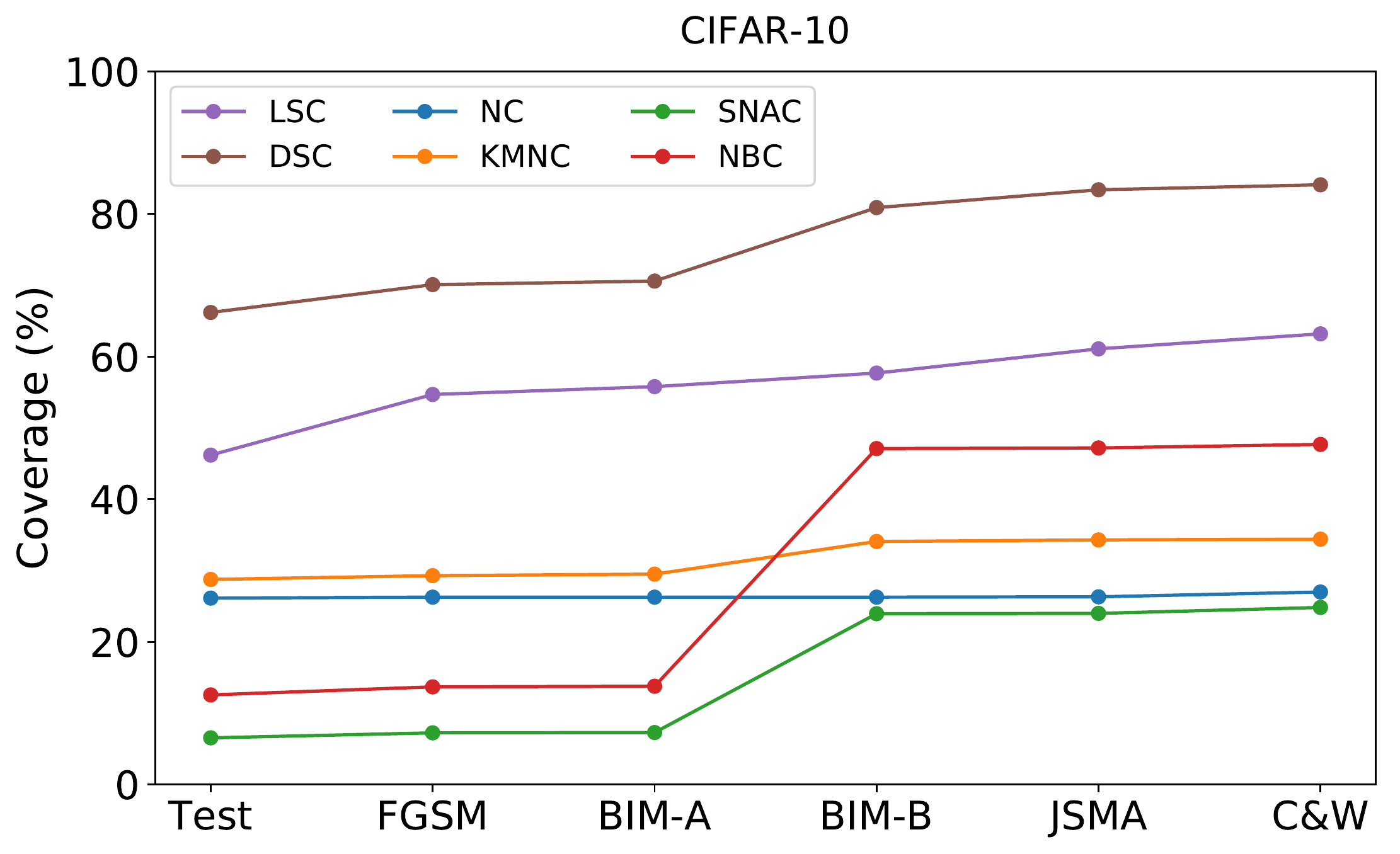}        
    \end{minipage}           
    \hspace{1.0em}
    \begin{minipage}[t]{.35\textwidth}
        \centering
        \includegraphics[width=\textwidth]{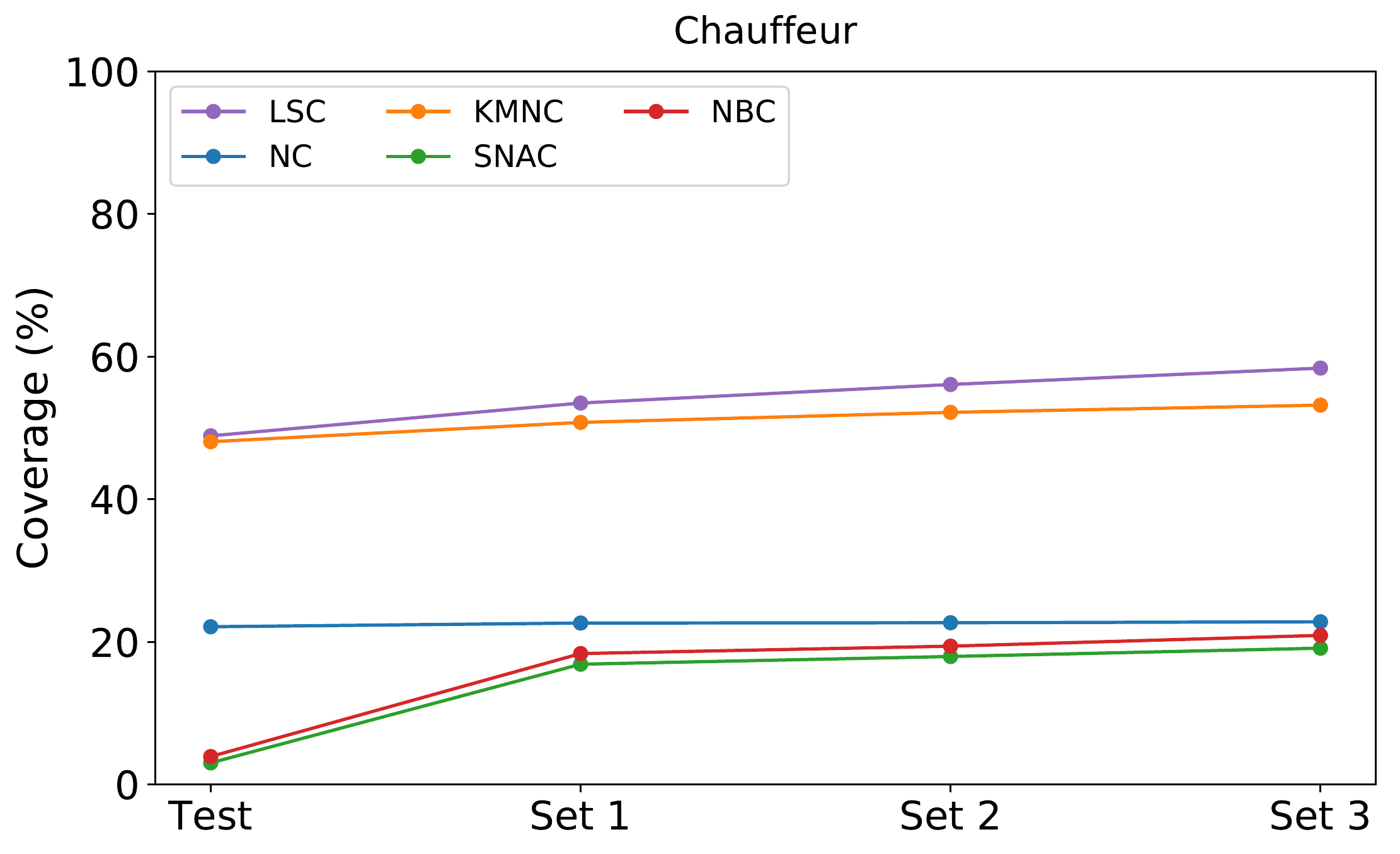}
    \end{minipage}   
    \caption{Visualisation of CIFAR-10 and Chauffeur in Table~\ref{table:RQ3_coverage}.    
    As additional sets of inputs ($x$-axis) are added to the original test set, 
    various coverage criteria ($y$-axis) increase.\label{fig:RQ3_coverage}}
\end{figure}

Table~\ref{table:RQ3_coverage} shows how different coverage criteria respond
to increasing diversity levels. Columns represent steps, at each of which more
inputs are added to the original test set. If the increase in coverage at a step
is less than 0.1 percentage point when compared to the previous step, the value
is underlined. The threshold of 0.1 percentage point is based on the finest
step change possible for LSC, DSC, as well as KMNC, as all three use bucketing 
with $k = 1,000$. We acknowledge that the threshold is arbitrary, and provide
it only as a supporting aid. Figure~\ref{fig:RQ3_coverage} presents 
visualisation of results from CIFAR-10 and Chauffeur. Note 
that DSC cannot be computed for these two DL systems, as they are not 
classifiers (see Section~\ref{sec:distance_sa}).

Overall, most of the studied criteria increase as additional inputs are added
at each step. The notable exception is NC, which plateaus against many steps.
This is in line with results in existing work~\cite{Ma2018ny}. There exists
an interplay between the type of added inputs and how different criteria
respond: SNAC, KMNC, and NBC show significant increases with the addition of 
BIM-B examples to CIFAR-10, but change little when C\&W inputs are added. 
However, only SNAC and NBC exhibit a similar increase with the addition of
input Set 1 for Chauffeur, while KMNC increases more steadily. Overall,
with the exception of NC, we answer 
RQ3 that \textbf{SC is correlated with other coverage criteria introduced so 
far}.

\begin{table}[ht] 
    \scalebox{0.75}{
    \begin{tabular}{clrrrrrr}
        \toprule
        DNN & Criteria   &   Test &   Step 1 & Step 2 &  Step 3 &  Step 4 & Step 5 \\
         &    &    &   + FGSM &  + BIM-A &   + BIM-B &   + JSMA &  + C\&W \\
        \midrule
        \multirow{6}{*}{MNIST} & LSC  & 29.50  & 34.90  & 37.10  & 56.30  & 61.90  & 62.00    \\
                               & DSC  & 46.00    & 56.10  & 65.00    & 67.20  & 70.90  & 72.30  \\
                               & NC   & 42.73 & \underline{42.73} & 43.03 & \underline{43.03} & \underline{43.03} & 45.45 \\
                               & KMNC & 68.42 & 70.96 & 72.24 & 75.82 & 77.31 & \underline{77.37} \\
                               & NBC  & 6.52  & 14.55 & 16.36 & 36.06 & 38.03 & 43.48 \\
                               & SNAC & 10.91 & 19.39 & \underline{19.39} & 53.33 & 57.27 & \underline{57.27} \\
        \midrule
        \multirow{6}{*}{CIFAR-10} & LSC  & 46.20  & 54.70  & 55.8  & 57.70  & 61.10  & 63.20  \\
                                  & DSC  & 66.20  & 70.10  & 70.6  & 80.90  & 83.40  & 84.10  \\
                                  & NC   & 26.15 & 26.28 & \underline{26.28} & \underline{26.28} & 26.33 & 27.01 \\
                                  & KMNC & 28.77 & 29.30  & 29.51 & 34.09 & 34.31 & 34.41 \\
                                  & NBC  & 6.56  & 7.26  & \underline{7.30}   & 23.96 & \underline{24.01} & 24.84 \\
                                  & SNAC & 12.58 & 13.71 & \underline{13.8}  & 47.11 & \underline{47.2}  & 47.70  \\
        \midrule
        \midrule
        DNN & Criteria   &   Test &  + SingleOcc & + MultiOcc & + Light & & \\
        \midrule
        \multirow{5}{*}{Dave-2} & LSC  & 30.00    & \underline{42.00}    & \underline{42.00}    & 76.00    & & \\
                                & NC   & 79.55 & 80.26 & 80.45 & 83.14 & & \\
                                & KMNC & 33.53 & 35.15 & 35.91 & 37.94 & & \\
                                & NBC  & 0.51  & 5.29  & \underline{5.32}  & 6.60   & & \\
                                & SNAC & 1.03  & 10.58 & \underline{10.64} & 13.21 & & \\
        \midrule        
        \midrule
        DNN &Criteria   &   Test &   + Set 1 &   + Set 2 &   + Set 3 \\
        \midrule
        \multirow{5}{*}{Chauffeur} & LSC  & 48.90  & 53.50  & 56.10  & 58.40  \\
                                   & NC   & 22.14 & 22.65 & \underline{22.70}  & 22.83 \\
                                   & KMNC & 48.08 & 50.79 & 52.20  & 53.21 \\
                                   & NBC  & 3.05  & 16.88 & 17.96 & 19.13 \\
                                   & SNAC & 3.93  & 18.37 & 19.41 & 20.93 \\
        \bottomrule
    \end{tabular}
    }
    \caption{Changes in various coverage criteria against increasing input diversity.
    We put additional inputs into the original test inputs %
    and observe changes in coverage values.\label{table:RQ3_coverage}}
\end{table}

\subsection{Retraining Guidance (RQ4)}
\label{sec:rq4}

Table~\ref{table:RQ4_retraining} shows the impact of SA-based guidance for 
retraining of MNIST, CIFAR-10, and Dave-2 models. The column $R$ from 
$\frac{1}{4}$ to $\frac{4}{4}$ represents the increasingly wider ranges
of SA from which the inputs for additional training are sampled; rows with 
$R = \emptyset$ show performance of the DL system before retraining.
Overall, 
there are 23 retraining configurations (2 SA types $\times$ 2 DL systems 
$\times$ 5 adversarial attack strategies, and 1 SA type $\times$ 1 DL system 
$\times$ three input synthesis methods), each of which is evaluated against
four SA ranges with 20 repetitions. Columns $\mu$ and $\sigma$ 
contain the mean and standard deviation of observed performance metric
(i.e., the highest accuracy for MNIST and CIFAR-10, the lowest MSE for Dave-2). 
The best performance is typeset in bold.

The full range, $\frac{4}{4}$, produces the best retraining performance for 13 
configurations, followed by $\frac{2}{4}$ (5 configurations), $\frac{3}{4}$ 
(3 configurations), and $\frac{1}{4}$ (3 configurations). Note that for the 
configuration of CIFAR-10 and BIM-B, both ranges $\frac{2}{4}$ and 
$\frac{2}{4}$ produces the same and the best retraining performance. The largest
improvement is observed when retraining MNIST against FGSM using DSA: the
accuracy of the $\frac{4}{4}$ range shows 77.5\% increase from that of 
$\frac{1}{4}$ (i.e., from 15.60\% to 28.69\%). While retraining MNIST against 
BIM-B using DSA shows even greater improvement (from 9.40\% to 40.94\%), 
we suspect this is an outlier
as the accuracy for ranges $\frac{1}{4}$ and $\frac{2}{4}$ are significantly
smaller when compared to other configurations.

While our observations are limited to the DL systems and input generation 
techniques studied here, we answer RQ4 that \textbf{SA can provide guidance 
for more effective retraining against adversarial examples based on our 
interpretation of the observed trend}. 

\begin{table}[ht]    

    \begin{minipage}[t]{0.5\textwidth}        
        \scalebox{0.68}{
        \begin{tabular}{c|c|c|rr|rr|rr|rr|rr}
            \toprule
            DNN & SA & $R$ &   \multicolumn{2}{c|}{FGSM} &   \multicolumn{2}{c|}{BIM-A} &    \multicolumn{2}{c|}{BIM-B} &   \multicolumn{2}{c|}{JSMA} & \multicolumn{2}{c}{C\&W} \\
            Model&  &   & $\mu$ & $\sigma$ & $\mu$ & $\sigma$ & $\mu$ & $\sigma$ & $\mu$ & $\sigma$ & $\mu$ & $\sigma$ \\
            \midrule
            \multirow{9}{*}{\rotatebox{90}{MNIST}} &  & $\emptyset$ & 11.65          & - & 9.38 & - & 9.38          & - & 18.88 & - & 8.92     & -     \\
            \cmidrule{2-13}
                                                 &   \multirow{4}{*}{LSA}                     & 1/4 & 25.81          & 1.95 & 95.14          & 0.69  & \textbf{41.00} &  0.01 & 72.67          & 3.09 & 92.51          & 0.51 \\
                                                 &                        & 2/4 & 28.45          & 2.91 & 95.71          & 0.41  & 40.98          &  0.12 & 75.03          & 2.68 & 92.55          & 0.67 \\
                                                 &                        & 3/4 & \textbf{29.66} & 3.63 & 95.87          & 0.98  & 40.97          &  0.10 & 75.48          & 2.60 & 92.41          & 1.03 \\
                                                 &                        & 4/4 & 23.70           & 4.98 & \textbf{95.90} & 0.79  & 40.93          &  0.18 & \textbf{77.37} & 1.75 & \textbf{92.56} & 0.77 \\
            \cmidrule{2-13} 
            & \multirow{4}{*}{DSA} & 1/4 & 15.60          & 2.12 & 93.67          & 3.42 & 9.90           & 1.05 & 74.56          & 2.62 & 12.80          & 0.96 \\
			&                        & 2/4 & 19.67          & 4.32 & \textbf{95.78} & 0.70 & 9.40           & 0.05 & 76.16          & 2.69 & 12.46          & 1.00 \\
			&                        & 3/4 & 26.37          & 6.15 & 95.37          & 0.93 & 40.81          & 0.22 & \textbf{78.01} & 1.87 & 12.37          & 1.14 \\
			&                        & 4/4 & \textbf{27.69} & 5.59 & 95.31          & 0.98 & \textbf{40.94} & 0.04 & 76.60           & 2.38 & \textbf{13.61} & 1.19 \\

            \midrule
            \midrule                             
            \multirow{9}{*}{\rotatebox{90}{CIFAR-10}} & & $\emptyset$ &   6.13 & - & 0.00 & - & 0.00 & - & 2.68 & - & 0.31 & -\\
            \cmidrule{2-13}
            & \multirow{4}{*}{LSA} & 1/4 & 11.07          & 1.20 & 32.34          & 1.70 & 0.59          & 1.76 & 32.80           & 2.05 & 34.38          & 2.83 \\
            &                        & 2/4 & \textbf{12.96} & 2.18 & 32.68          & 2.07 & \textbf{0.89} & 2.10 & 33.84          & 2.52 & 42.99          & 2.78 \\
            &                        & 3/4 & 12.79          & 2.17 & 32.14          & 2.40 & \textbf{0.89} & 2.10 & 35.81          & 2.81 & 45.58          & 2.23 \\
            &                        & 4/4 & 12.53          & 1.19 & \textbf{32.79} & 2.29 & 0.60          & 1.76 & \textbf{35.83} & 2.54 & \textbf{45.74} & 2.04 \\                                     
            \cmidrule{2-13} 
            & \multirow{4}{*}{DSA} & 1/4 & \textbf{14.86} & 2.16 & 25.94          & 2.99 & 0.01          & 0.00 & 34.92          & 2.01 & 44.21          & 2.02 \\
            &                        & 2/4 & 14.64          & 1.95 & 29.59          & 3.52 & 0.01          & 0.00 & 34.49          & 1.89 & 44.79          & 2.32 \\
            &                        & 3/4 & 13.81          & 1.85 & 31.93          & 2.77 & 0.01          & 0.00 & 35.61          & 2.40 & 46.16          & 2.45 \\
            &                        & 4/4 & 13.12          & 1.41 & \textbf{32.17} & 2.36 & \textbf{0.60} & 1.76 & \textbf{37.32} & 1.58 & \textbf{46.21} & 2.72 \\
            \bottomrule
        \end{tabular} 
        }
        \subcaption{MNIST and CIFAR-10}               
        \vspace{1.0em}
    \end{minipage}        
    
    \begin{minipage}[t]{0.5\textwidth}
        \centering
        \scalebox{0.8}{
        \begin{tabular}{c|c|c|rr|rr|rr}
            \toprule
            DNN  & SA & $R$   &  \multicolumn{2}{c|}{SingleOcc} & \multicolumn{2}{c|}{MultiOcc} &  \multicolumn{2}{c}{Light} \\
            Model &  &   & $\mu$ & $\sigma$ & $\mu$ & $\sigma$ & $\mu$ & $\sigma$ \\
            \midrule
            \multirow{5}{*}{Dave-2} &  & $\emptyset$ & 0.4212     & - & 0.0964 & - & 0.3822 & - \\
            \cmidrule{2-9}
                                    & \multirow{4}{*}{LSA} & 1/4 & 0.0586         & 0.0142 & \textbf{0.0539} & 0.0003 & 0.0573 & 0.0057 \\
                                    &                         & 2/4 & \textbf{0.0540} & 0.0012 & 0.0562          & 0.0060 & \textbf{0.0560}  & 0.0042 \\
                                    &                         & 3/4 & 0.0554         & 0.0041 & 0.0544          & 0.0009 & 0.0570  & 0.0133 \\
                                    &                         & 4/4 & 0.0553         & 0.0028 & 0.0561          & 0.0042 & 0.0601 & 0.0111 \\
            \bottomrule
        \end{tabular}
        }
        \subcaption{Dave-2}
    \end{minipage}
    
    \caption{Retraining guided by SA: we sample 100 inputs from four increasingly wider ranges of SA:
         $[0, \frac{U}{4}]$, $[0, \frac{2U}{4}]$, $[0, \frac{3U}{4}]$, and $[0, U]$, 
         and retrain for five additional epochs using the samples as the training data,
         and measure the accuracy and MSE against the entire adversarial and synthetic inputs.
         Sampling from wider ranges improves the retraining accuracy.\label{table:RQ4_retraining}}
\end{table}

\section{Threats to Validity}
\label{sec:threats}

The primary threat to internal validity of this study is the correctness of
implementation of the studied DL systems, as well as the computation of SA
values. We have used publicly available architectures and pre-trained models 
as our subjects to avoid incorrect implementation. SA computation depends on 
a widely used computation library, \texttt{SciPy}, which has stood the public 
scrutiny. Threats to external validity mostly concerns the number of the models
and input generation techniques we study here. It is possible that 
\name is less effective against other DL systems. While we believe the 
core principle of measuring input surprise is universally applicable, only 
further experimentations can reduce this particular risk. Finally, threats 
to construct validity asks whether we are measuring the correct factors to 
draw our conclusion. For all studied DL systems, activation traces are 
immediate artefacts of their executions and the meaning of output accuracy 
is well established, minimising the risk of this threat.

\section{Related Work}
\label{sec:relatedwork}

Adversarial examples pose significant threats to the performance of DL 
systems~\cite{Carlini2017aa}. There are existing work in the machine learning 
community on detection of such inputs. Feinman et al.~\cite{Feinman2017aa} 
first introduced the KDE as a means of similarity measurement, with the aim 
of detecting adversarial examples. 
\name improves upon the existing work by a number of different ways. 
First, we generalise the concept of Surprise Adequacy (SA) and introduce 
Distance-based SA. Second, our evaluation is in the context of 
DL system testing. Third, our evaluation of \name includes more complicated
and practical DL systems, as well as testing techniques such as
DeepXplore and DeepTest. Finally, we show that the choice of neurons has 
limited impact on LSA.

A range of 
techniques has been recently proposed to test and verify DL systems. The 
existing techniques are largely based on two assumptions. The first assumption 
is a variation of metamorphic testing~\cite{Chen:2004th,Murphy:2009fy,
Yoo:2010vn}. Suppose a DL system $N$ produces an output $o$ when given $i$ as 
the input, i.e., $N(i) = o$. Then we expect $N(i') \simeq o$ when $i' \simeq i$
. Huang et al.~\cite{Huang2017kx} proposed a verification technique that can 
automatically generate counter-examples that violate this assumption. Pei et 
al. introduced DeepXplore~\cite{Pei2017qy}, a white-box technique that 
generates test inputs that cause disagreement among a set of DL systems, 
i.e., $N_m(i) \neq N_n(i)$ for independently trained DL systems $N_m$ 
and $N_n$. Tian et al. presented DeepTest, whose metamorphic relations include 
both simple geometric perturbations as well as realistic weather 
effects~\cite{Tian2018zn}. The second assumption is that the more diverse a 
set of input is, the more effective it will be for testing and validating DL 
systems. Pei et al. proposed Neuron Coverage (NC), which measures the ratio of 
neurons whose activation values are above a predefined 
threshold~\cite{Pei2017qy}. It has been shown that adding test 
inputs that violate the first assumption increases the diversity measured 
through NC. Similarly, DeepGauge introduced a set of multi-granularity coverage 
criteria that are thought to reflect behaviours of DL systems in finer
granularity~\cite{Ma2018ny}. While these criteria capture input
diversity, all of them are essentially count of neurons unlike SA, and 
therefore cannot be directly linked to behaviours of DL systems. We show 
that SA is closely related to the behaviours by training accurate 
adversarial example classifiers based on SA.

Apart from coverage criteria, other concepts in traditional software testing 
have been reformulated and applied to testing of DL systems. 
Ma et al. proposed DeepCT, which views ranges of neuron activation values
as parameter choices and applies Combinatorial Interaction Testing (CIT) to
measure interaction coverage~\cite{ma2018combinatorial}. SC is different
from DeepCT as \name aims to quantify the amount of surprise, rather than
simply to detect surprise via increase in coverage.
DeepMutation applies the principle of mutation testing to DL systems by
mutating training data, test data, as well as the DL system itself,
based on source and model level mutation operators~\cite{ma2018deepmutation}.

\section{Conclusion}
\label{sec:conclusion}

We propose \name, a surprise adequacy framework for DL systems that can 
quantitatively measure relative surprise of each input with respect to the 
training data, which we call Surprise Adequacy (SA). Using SA, we also develop
Surprise Coverage (SC), which measures the coverage of discretised input
surprise ranges, rather than the count of neurons with specific activation 
traits. Our empirical evaluation shows that SA and SC can capture the surprise 
of inputs accurately and are good indicators of how DL systems will react to 
unknown inputs. SA is correlated with how difficult a DL system finds an input,
and can be used to accurately classify adversarial examples. SC can be used to
guide selection of inputs for more effective retraining of DL systems for 
adversarial examples as well as inputs synthesised by DeepXplore. 

\bibliography{newref}
\bibliographystyle{plain}

\end{document}